\documentclass[a4paper,fleqn,usenatbib]{mnras}

\usepackage[T1]{fontenc}
\usepackage{ae,aecompl}

\usepackage{graphicx}	
\usepackage{amsmath}	
\usepackage{amssymb}	
\usepackage{times}

\usepackage{hyperref}
\hypersetup{
  citecolor=cyan,      
}

\usepackage{graphicx}
\usepackage{color}
\definecolor{orange}{rgb}{1,0.5,0}
\definecolor{purple}{rgb}{0.6, 0.2, 0.8}

\newcommand{\ie}{i.e.,~}
\newcommand{\eg}{e.g.,~}

\usepackage{amsmath}
\usepackage{amssymb}
\usepackage{mathtools}

\title[Accretion disks formed in BH--NS mergers]{
  On accretion disks formed in MHD simulations of black-hole--neutron
  star mergers with accurate microphysics}

\author[Elias R. Most et al.]{ Elias R. Most$^{1,2,3}$
\thanks{Corresponding author: emost@princeton.edu}, L. Jens
Papenfort$^{4}$,
Samuel D. Tootle$^{4}$, Luciano Rezzolla $^{4,5,6}$\\ 
$^{1}${Princeton Center for Theoretical Science, Princeton University,
Princeton, NJ 08544, USA}\\
$^{2}${Princeton Gravity Initiative, Princeton University, Princeton, NJ
08544, USA}\\
$^{3}${School of Natural Sciences, Institute for Advanced Study, Princeton,
NJ 08540, USA}\\
$^{4}${Institut f\"ur Theoretische Physik, Goethe Universit\"at,
Max-von-Laue-Str. 1, 60438 Frankfurt am Main, Germany}\\
$^5${School of Mathematics, Trinity College, Dublin 2, Ireland}\\
$^6${Frankfurt Institute for Advanced Studies, Ruth-Moufang-Str. 1, 60438
Frankfurt am Main, Germany}
}

\date{Accepted XXX. Received YYY; in original form ZZZ}

\pubyear{2021}

\begin{document}
\label{firstpage}
\maketitle
\begin{abstract}
  Remnant accretion disks formed in compact object mergers are an
  important ingredient in the understanding of electromagnetic afterglows
  of multi-messenger gravitational-wave events. Due to magnetically and
  neutrino driven winds, a significant fraction of the disk mass will
  eventually become unbound and undergo r-process nucleosynthesis. While
  this process has been studied in some detail, previous studies have
  typically used approximate initial conditions for the accretion disks,
  or started from purely hydrodynamical simulations. In this work, we
  analyse the properties of accretion disks formed from near equal-mass
  black hole-neutron star mergers simulated in general-relativistic
  magnetohydrodynamics in dynamical spacetimes with an accurate
  microphysical description. The post-merger systems were evolved until
  $120\, {\rm ms}$ for different finite-temperature equations of state
  and black-hole spins. We present a detailed analysis of the fluid
  properties and of the magnetic-field topology. In particular, we
  provide analytic fits of the magnetic-field strength and specific
  entropy as a function of the rest-mass density, which can be used for
  the construction of equilibrium disk models. Finally, we evolve one of
  the systems for a total of $350\, \rm ms$ after merger and study the
  prospect for eventual jet launching. While our simulations do not reach
  this stage, we find clear evidence of continued funnel magnetization
  and clearing, a prerequisite for any jet-launching mechanism.
\end{abstract}


\section{Introduction} 
\label{sec:intro}

\begin{table*}
  \centering
  \begin{tabular}{{|l|c|c|c|c|c|c|c|c|c|}}

    \hline
    \hline
    &
    $m_{\rm BH}\left[M_\odot\right]$  &
    $m_{\rm NS}\left[M_\odot\right]$ &
     $m_{b} \left[M_\odot\right]$ 
     &$q$ &
    $\chi_{\rm BH}$ & 
    $\tilde{\chi}$ &
  $M_b^{\rm fin} \left[M_\odot\right]$&
    ${\chi_{\rm BH}^{\rm fin}}$&
    ${\rm EOS}$\\
    \hline
    \hline
    \texttt{TNT.chit.0.00} & $2.20$ & $1.40$ & $1.55$ & $0.636$ & $0.00$ & $0.00$  & $0.013$  &$0.73$  &{\texttt{TNTYST}} \\
    \texttt{TNT.chit.0.15} & $2.24$ & $1.36$ & $1.50$ & $0.608$ & $0.24$ & $0.15$  & $0.057$  &$0.79$  &{\texttt{TNTYST}} \\
    \texttt{TNT.chit.0.35} & $2.42$ & $1.18$ & $1.28$ & $0.486$ & $0.52$ & $0.35$  & $0.170$  &$0.83$  &{\texttt{TNTYST}} \\
       \hline                                                                                
    \texttt{BHBLP.BH.chit.0.00} & $2.10$  & $1.50$  & $1.65$ &$0.636$ & $0.00$ & $0.00$  & $0.033$  &$0.77$  &{\texttt{BHB}$\Lambda\Phi$} \\
    \texttt{BHBLP.BH.chit.0.15} & $2.14$  & $1.46$  & $1.60$ &$0.608$ & $0.24$ & $0.15$  & $0.073$  &$0.82$  &{\texttt{BHB}$\Lambda\Phi$} \\
    \texttt{BHBLP.BH.chit.0.35} & $2.33$  & $1.27$  & $1.37$ &$0.543$ & $0.54$ & $0.35$  & $0.124$  &$0.86$  &{\texttt{BHB}$\Lambda\Phi$} \\
       \hline                                                                                
       \hline                                                                                
  \end{tabular}
  \caption{Summary of the properties of the initial binaries. The columns
    list: the gravitational component masses in isolation, $m_{\rm BH}$
    and $m_{\rm NS}$, the baryon mass $m_{b}$ of the secondary, the mass
    ratio $q=m_{\rm NS}/m_{\rm BH}$, the dimensionless spin $\chi_{\rm
      BH}$ of the BH (\textit{primary}), the effective spin $\tilde{\chi}
    := \chi_{\rm 1}/ (1+q)$ of the binary and the $\rm EOS$ describing
    the nuclear matter. All binaries have a total mass $M_{_{\rm
        ADM}}=3.6\, M_\odot$ and are at an initial separation of
    $45\,{\rm km}$. The NS (\textit{secondary}) is always nonrotating,
    $\chi_{\rm NS}=0$. We also state the final, post-merger BH spins
    $\chi_{\rm BH}^{\rm fin}$ and disk masses $M_b^{\rm fin}$ first
    reported in \citet{Most2020e}.}
    \label{tab:initial}
\end{table*}

Recently, LIGO has announced early results of the third observing run,
indicating the potential detection of several black hole (BH) -- neutron
star (NS) systems. Two of them, S200105ae and S200115j, have recently
been studied by optical follow-up observations \citep{Anand2021}, albeit
no kilonova afterglow has so far been detected. Since the kilonova of
these systems would be mainly driven by secular disk mass ejecta, a
potential non-detection can set tight constraints on the allowed range of
parameters. In particular, retaining a sufficient disk mass after the
merger of a small mass-ratio system also requires high BH spin
\citep{Foucart2018b}. Therefore, a non-detection in conjunction with the
inferred binary parameters from the inspiral allows to probe the BH spin
\citep{Anand2021,Raaijmakers2021}, and for low-mass BHs also the equation
of state (EOS) \citep{Fragione2021}. Several works have also further investigated
the prospects for constraining the EOS with BH-NS gravitational-wave
events (\eg \citet{Pannarale2011, Maselli2013b, Lackey2013}). While an
initial estimate of the amount of mass ejection can easily be done, a
more accurate calculation requires more precise knowledge of the amount
of unbound disk mass, its nuclear composition, velocity and temperature
distributions. In turn, this necessitates a careful investigation of the
initial accretion disks from which these mass outflows originate.

The merger and early post-merger of BH-NS systems have been explored in
great detail, placing an emphasis on the disk formation, mass ejection
and gravitational-wave emission \citep{Shibata06d, ShibataUryu:2007,
  ShibataTaniguchi2008, Etienne08, Etienne:2008re, Kyutoku2010,
  Pannarale2010, Foucart2010, Kyutoku2011, Foucart2011, Foucart2012,
  Foucart2013b, Kyutoku2015}. Whereas those simulations have focussed on
quasi-circular binaries most relevant for gravitational-wave detections,
some simulations have also investigated the merger of eccentric
encounters \citep{East2015}. More recent studies have also included
finite-temperature equations of state and neutrino transport
\citep{Foucart2013a,Foucart2014,Foucart2015a,Foucart2017b,Kyutoku2018},
allowing them to investigate the nuclear composition of the mass ejecta.
Based on such numerical simulations, it has also been possible to
accurately predict the masses of the disks formed in these mergers
\citep{Foucart2012,Foucart2018b}. Although simulations have mainly been
performed for mass ratios $q:= m_2/m_1 < 1/4$ -- where $m_{1,2}$ are the
masses of the binary components -- a few studies have been conducted for
systems in the near equal-mass regime
\citep{Hinderer2018,Foucart2019,Hayashi2020}, which is also the focus of
this work. While some of the remnant accretion disks formed in all of
these simulations have been studied with superimposed magnetic fields to
understand their long-term evolution \citep{Fernandez2017,Nouri2018},
relatively few general-relativistic BH-NS merger simulations with
magnetic fields initially confined to the NS have been
conducted. Practically all of them have used polytropic equations of
state \citep{Chawla:2010sw, Etienne2012, Etienne2012b, Paschalidis2014,
  Kiuchi2015, Wan2017, Ruiz2018}. To the best of our knowledge, this work
presents the first study to self-consistently investigate the merger and
post-merger of BH-NS systems with initial NS magnetic fields,
finite-temperature equations of state and neutrino leakage, where the
latter has been found to reasonably approximate the evolution of the
nuclear composition in the cold $\left( T<10\, \rm MeV \right)$ accretion
disks present in these systems \citep{Kyutoku2018}.

More specifically, we study the post-merger formation of an accretion
disk in nearl equal-mass BH-NS mergers for two finite-temperature EOS
and a magnetic field initially confined to the NS. The early post-merger
evolution and mass ejection of these systems has been presented in
\citet{Most2020e}. The follow-up simulations presented here highlight
the early magnetic field and composition evolution until $\sim 120\,\rm
ms$, and provide a detailed account of the properties of the accretion
disk formed in the merger and along its subsequent evolution. Finally,
we evolve one of the configurations for up to $350\, \rm ms$ and comment
on the prospects of jet launching from such systems.

\section{Methods}

In this section we provide a short overview of numerical methods and the
initial conditions used in this study.

We model the initial BH-NS systems as having irrotational NSs and
spinning BHs on quasi-circular orbits
\citep{Grandclement06,Papenfort2021}. The initial models are valid
solutions of the constraint sector of the Einstein equations constructed
using the conformally flat XCTS formalism. The excision boundary
conditions on the BH include a Neumann boundary condition on the lapse
and a tangential shift condition to control the quasi-local spin of the
BH \citep{Caudill:2006hw}. The interior of the BH is initially
regularized by extrapolating the solution using an eighth-order
Lagrangian polynomial along the radial direction
\citep{Etienne2007a,Etienne:2008re}. The NS is described by either of two
EOSs, \texttt{TNTYST} \citep{Togashi2017} or \texttt{BHB$\Lambda\Phi$}
\citep{Banik2014} in line with multi-messenger constraints on the NS
maximum mass \citep{Margalit2017, Rezzolla2017, Ruiz2017, Shibata2019,
  Nathanail2021} and radius \citep{Annala2018, Most2018, De2018,
  Abbott2018b,Raithel2018}. The initial NSs are endowed with an internal dipole field
via the vector potential $A_\varphi = \varpi^2 A_0 \max\left( p-0.04\,
p_\mathbf{max} ,0 \right)^2$, commonly used in these types of simulations
\citep{Liu:2008xy, Giacomazzo:2010, Etienne2012, Kiuchi2015}, where
$p_\mathbf{max}$ refers to the maximum pressure in the NS. The
coefficient, $A_0$, is chosen such that the maximum field strength in the
center of the star corresponds to $\simeq 10^{14}\, \rm G$. A summary of
the initial conditions is given in Tab. \ref{tab:initial}.

In general, it is both interesting and important to understand the
behaviour of a variety of disks in terms of disk masses, compositions and
magnetic-field topologies, in order to get a good coverage of the large
parameter space. Since we are interested in studying near equal-mass
systems consistent also with very massive NS binaries, we focus on
systems along the stability line (in terms of spins $\chi$) of the most
massive NSs \citep{Most2020c}, as presented in \citet{Most2020e}. This
assumption correlates the BH mass $M_{\rm BH}$ with the BH spin
$\chi_{\rm BH}$, leading to a parametrization only dependent on the
maximum mass $M_{\rm TOV}$ of a nonrotating NS \citep{Most2020c}. More
details on this construction can be found in \citet{Most2020c}.

\begin{figure*}
  \centering
  \includegraphics[width=\textwidth]{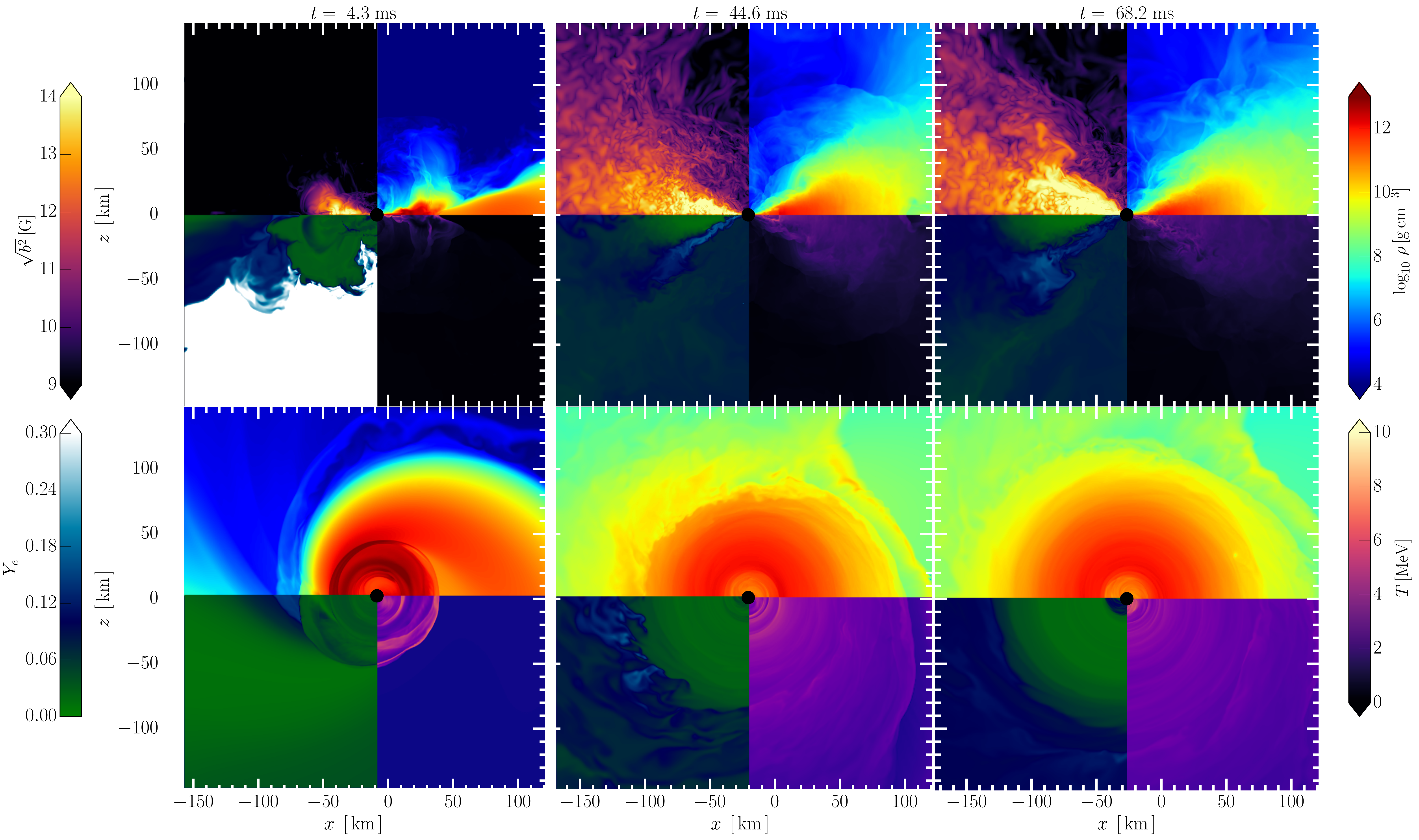}
  \caption{Model \texttt{TNT.chit.0.35}. Shown are the co-moving
    magnetic energy density $b^2$, the rest-mass density $\rho$, the
    electron fraction $Y_e$ and the local fluid temperature $T$. The
    different rows correspond to meridional {\it (Top)} and equatorial
    {\it (Bottom)} views of the accretion disk around the BH. The
    columns correspond to different times after merger, starting from the
    early formation of the disk after the star has been tidally disrupted
    (left column). The centre and right columns then refer to two times
    where the disk grows due to fall-back accretion of the bound tidal
    arm, and the onset of a steady accretion flow.}
  \label{fig:gw190425}
\end{figure*}

\subsection{Numerical methods}

\begin{figure*}
  \centering
  \includegraphics[width=\textwidth, trim=0. 2.4cm 0. 0. , clip]{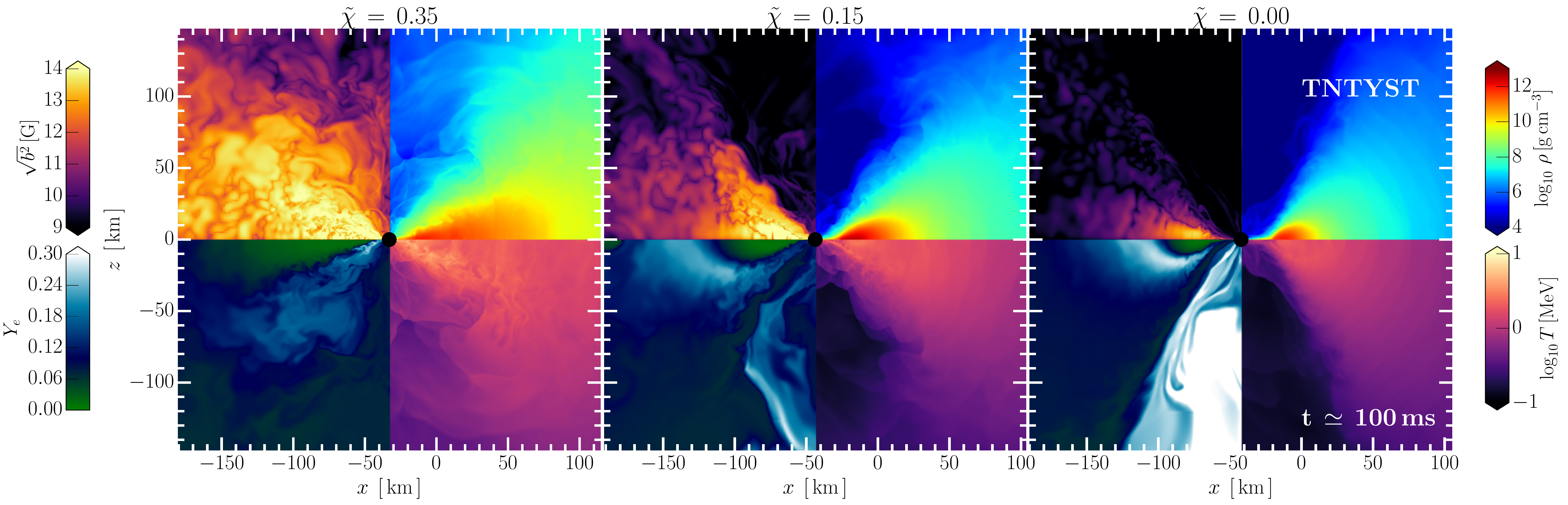}
  \includegraphics[width=\textwidth, trim=0. 0. 0. 1.cm, clip]{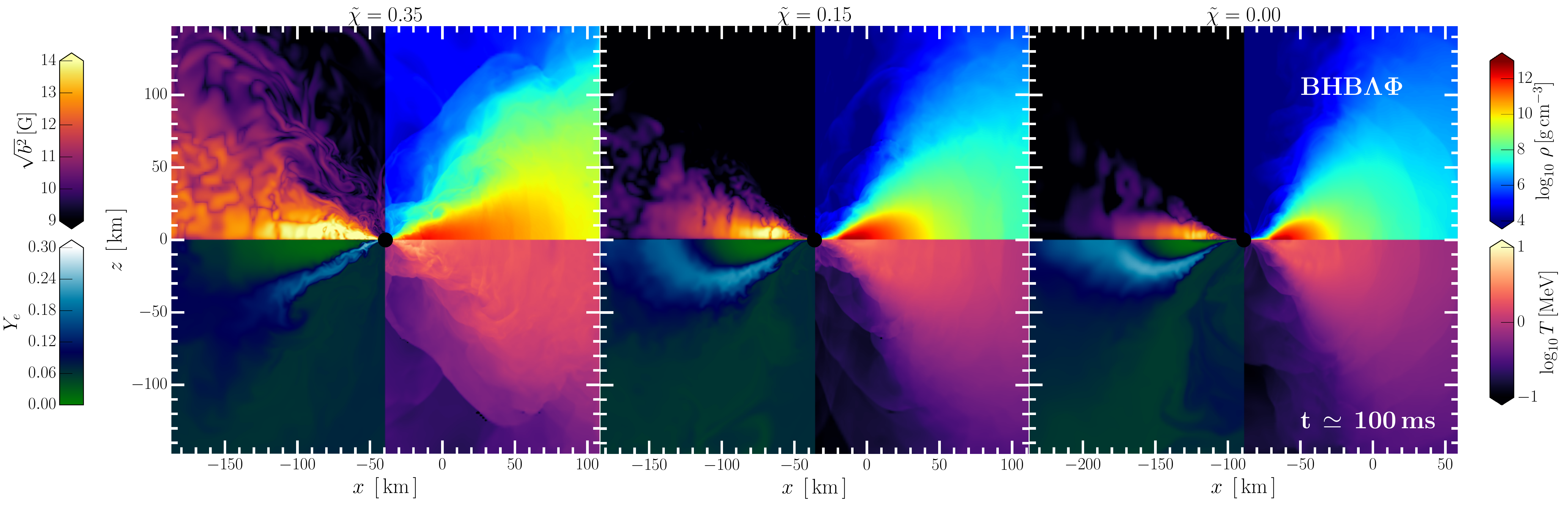}
  \caption{Shown are the evolution of the co-moving magnetic energy
    density $b^2$, the rest-mass density $\rho$, the electron fraction
    $Y_e$ and the local fluid temperature $T$ in the meridional
    plane. The top row shows models computed using the \texttt{TNTYST}
    EOS, while the bottom row the \texttt{BHB}$\Lambda\Phi$ EOS. All
    results are shown at $\simeq 100\, \rm ms$ after merger.
  } \label{fig:xz_tntyst}
\end{figure*}

In order to model the dynamical evolution of the BH-NS system, we solve
the general-relativistic ideal magnetohydrodynamics (GRMHD) equations
together with the Einstein field equations (EFE),
\begin{align} 
  \nabla_\mu T^{\mu\nu} &= -\mathcal{Q}^\nu\,, \\
  R_{\mu\nu} &= 8 \pi \left( T_{\mu\nu} - \frac{1}{2} g_{\lambda\kappa} T^{\kappa\lambda}
g_{\mu\nu} \right) \nonumber
  \\&+ 2 \nabla_{\left(\mu \right.} Z_{\left. \nu \right)}
  + \kappa_1 \left[ 2 n_{\left( \mu \right.} Z_{\left. \nu\right)} -
  n_\alpha Z^\alpha g_{\mu \nu} \right]\,, 
  \label{eqn:Einstein}
\end{align} 
where $g_{\mu\nu}$ is the four-dimensional Lorentzian spacetime metric,
$R_{\mu\nu}$ the corresponding Ricci tensor and $T_{\mu\nu}$ the energy
momentum tensor describing the NS matter and the magnetic
fields.

The source term $\mathcal{Q}^\nu$ represents the energy and momentum loss
due to weak interactions. Using the unit normal vector $n_\mu$ of the 3+1
slicing of the spacetime \citep{Gourgoulhon2012} and the Z-vector $Z_\mu$
within the Z4 system \citep{Bona:2003fj}, the EFE are written as a system
that allows for the propagation of numerical constraint violations of the
Einstein system \citep{Gundlach2005:constraint-damping}. We solve the EFE
using the Z4c formulation \citep{Hilditch2013,Bernuzzi:2009ex}, which is
a conformal variant of the Z4 system \citep{Bona:2003fj} (see also
\citealt{Alic:2011a}). Different from \citet{Weyhausen:2011cg}, we find
that simulations of BH-NS binaries employing the excision formalism on
the initial data require additional damping, $\kappa_1=0.07$, whereas
larger damping leads to instabilities of the spacetime evolution. In
addition, we find it beneficial to remove the advection part in the shift
condition, which is then given by \citep{Alcubierre2003,Etienne2007b},
\begin{align}
  \partial_t \beta^i &= B^i\,, \\
  \partial_t B^i &= \frac{3}{4} \bar{\Gamma}^i - \eta B^i\,,
  \label{eqn:gamma_driver}
\end{align}
with damping parameter $\eta = 1.4$.

The ideal-GRMHD equations \citep{Duez05MHD0,Shibata05b,Giacomazzo:2007ti}
are supplemented by an evolution equation for the magnetic vector
potential in the ideal-MHD limit \citep{DelZanna2003,Etienne:2010ui}. We
additionally impose the Lorenz gauge for the vector potential
\citep{Etienne2012a}. Neutrino losses are incorporated using a simplified
leakage prescription \citep{Ruffert96b,Rosswog:2003b,Galeazzi2013}, that
is appropriate for the low temperatures reached in the tidal disruption
of a NS \citep{Deaton2013,Kyutoku2018}.

These equations are solved using the \texttt{Frankfurt\-/IllinoisGRMHD}
code (\texttt{FIL}) \citep{Most2019b,Most2018b}. Although \texttt{FIL} is
derived from the \texttt{IllinoisGRMHD} code \citep{Etienne2015}, it
makes use of a fully fourth-order conservative finite-difference
algorithm to discretize the hydrodynamical and electromagnetic flux terms
\citep{DelZanna2007}. Furthermore, it provides routines to use tabulated
finite-temperature EOSs and can evolve the electron fraction $Y_e$. In
addition, \texttt{FIL} solves the Z4c system using fourth-order accurate
upwinded finite-differences \citep{Zlochower2005:fourth-order}. Details
on the implementation and accuracy of the code can be found in
\citet{Most2019b}.

\texttt{FIL} is built on top of the Einstein Toolkit
\citep{loeffler_2011_et,ET2019}. As such, \texttt{FIL} uses a fixed-mesh
box-in-box refinement provided by \texttt{Carpet}
\citep{Schnetter-etal-03b}. Specifically, we use nine nested Cartesian
boxes each at doubling resolution. The outer domain extends to $\simeq
6000\,\rm km$ in each direction and the initial compact objects are
covered by the two finest domains with a size of $17.7\, \rm km$ and a
resolution of $\simeq 215\, \rm m$. Additionally, we impose reflection
symmetry along the vertical $z-$direction.

\section{Results}

In this work, we study the merger and post-merger evolution of near
equal-mass BH-NS binaries. Before turning to the properties of the
accretion disks formed in such mergers, we first provide a very brief
overview of their formation. We do this by considering the fiducial
system \texttt{TNT.chit.0.35}. In order to illustrate the disk formation
process, we begin by summarizing the dynamical formation of the disk in
Fig. \ref{fig:gw190425}, which reports the co-moving magnetic energy
density $b^2$, the rest-mass density $\rho$, the electron fraction $Y_e$
and the local fluid temperature $T$. The different rows correspond to
meridional (top panels) and equatorial (bottom panels) views of the
accretion disk around the BH, while the different columns correspond to
different times after the merger. The general dynamics of this process
have been studied extensively in purely hydrodynamical simulations
\citep{Etienne:2008re,Kyutoku2011,Foucart2011}. In order for a massive
disk to form during and after merger, tidal disruption has to occur
outside of the innermost stable circular orbit (ISCO) of the BH
\citep{Pannarale2010, ShibataTaniguchilrr-2011-6}. Starting from the left
panel, we can see that shortly after tidal disruption, an initial
accretion disk begins to form around the BH. Originating from the cold NS
matter, the initial disk is very neutron rich ($Y_e < 0.05$), but already
reaches temperatures $T \lesssim 10\, \rm MeV$. The disk quickly grows in
mass and size due to fall-back accretion from the tidal arm (middle
column), begins to circularize and a steady accretion flow develops over
time. As expected, this happens on the dynamical timescales of the disks,
which are proportional to the disk mass $M_{\rm disk}^b$, so that the
lightest disks circularize first. Initially, the pure neutron matter is
far out of beta-equilibrium under these conditions and will rapidly
re-equilibrate via beta decay of neutrons, leading to an increasing
protonization especially of the low-density parts of the disk. At the
same time, the magnetic-field strength is increasing throughout the disk,
exceeding $10^{14}\, \rm G$ locally. More details on the magnetic-field
evolution will be given in Sec. \ref{sec:mag}. Finally, after more than
$50\, \rm ms$ past merger, the disk has settled into an initial
quasi-equilibrium, consisting of a very neutron-rich disk, probing
rest-mass densities $\lesssim 10^{11}\, \rm g\,cm^{-3}$. A disk formed by
this process will then set the initial conditions for the long-term
evolution in terms of the accretion flow and mass ejection
\citep{Fernandez2015b,Fernandez2017}.

\begin{figure*}
  \centering
  \includegraphics[width=0.9\textwidth]{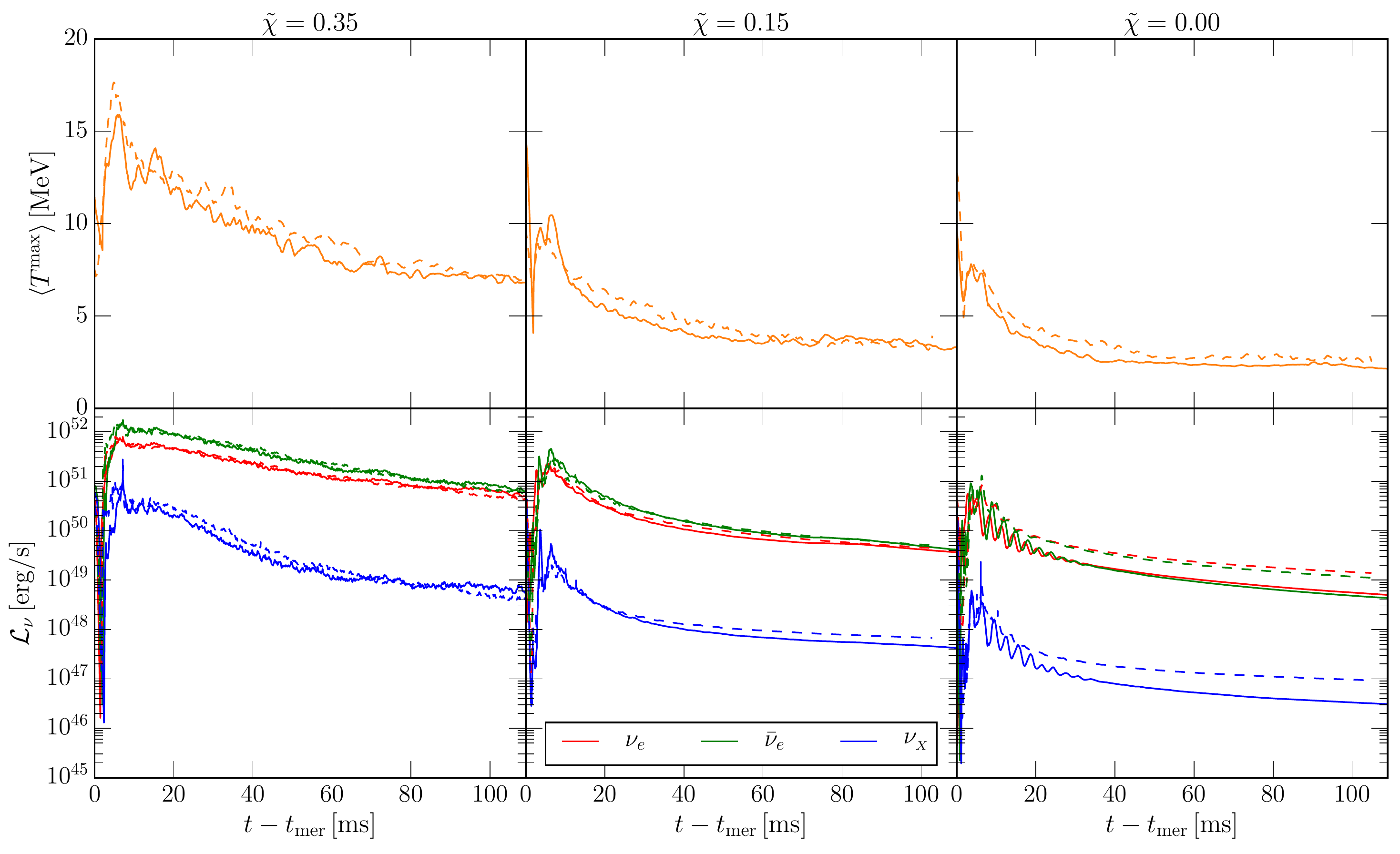}
  \caption{Temperature evolution and neutrino emission of the disk. {\it
      (Top)} time-averaged maximum temperature $\langle T^{\max}\rangle$.
    {\it (Bottom)} neutrino luminosity $\mathcal{L}_\nu$. Results are
    shown for simulations with the \texttt{TNTYST} (solid) and
    \texttt{BHB}$\Lambda\Phi$ EOSs. The columns represent the different
    models in terms of their effective spin $\tilde{\chi}$}
  \label{fig:hydro_prop_tntyst}
\end{figure*}

\subsection{Disk properties}


One of the most important observables of such a gravitational-wave event
would be the associated optical counterparts, in particular the kilonova
afterglow \citep[see][for a review]{Metzger2017}. In the case of a BH--NS
merger with a massive remnant accretion disk, this will be caused
primarily by secular (essentially magnetically and neutrino driven) disk
mass ejection \citep{Fernandez2013,Fernandez2015b,Siegel2017}. In
addition, dynamical mass ejection will also lead to an early red kilonova
component
\citep{Kyutoku2013,Kyutoku2015,Foucart2013b,Kawaguchi2016}. Previous
simulations of realistic (\eg \citet{Fernandez2017,Nouri2018}) or
idealised (\eg \citet{Siegel2017,Fernandez2018}) remnant disks have shown
that a large fraction of the disk material will become
unbound. Therefore, it is important to understand the initial structure
of realistic disks formed during merger by the tidal disruption of the
NS.

\subsubsection{General observations}

In this section we will focus on the composition of the disk formed after
the merger and will highlight part of its evolution. As discussed in in
the previous section using Fig. \ref{fig:gw190425}, the disk is formed by
the tidal disruption of the NS at merger. The neutron-rich debris then
forms an accretion disk, which will become quasi-stationary once the
fall-back accretion of matter from the bound part of the tidal tail has
ceased. Generically, we find that this happens after
roughly $ 50-70\, \rm ms$.

Having outlined the general stages of the disk formation process, we now
present the resulting disks for three different effective spins
$\tilde{\chi} := \chi_{\rm 1}/ (1+q) = \left[ 0.00\,, 0.15\,,0.35
  \right]$ of the binary, performed with both the \texttt{TNTYST} and
\texttt{BHB}$\Lambda\Phi$ EOS. In Fig. \ref{fig:xz_tntyst} the rest-mass
density $\rho$, electron fraction $Y_e$ and temperature $T$ are shown on
the equatorial plane for the three effective spins at a time $t\approx
100\, \rm ms$ after merger.

We begin by discussing the evolution of the \texttt{TNTYST} systems (top
row), although all conclusions will essentially also hold for the
\texttt{BHB}$\Lambda\Phi$ EOS, as can be seen from the bottom row of
Fig. \ref{fig:xz_tntyst}. This is because the smaller compactness $C= M/R$
of the initial NS for the \texttt{BHB}$\Lambda\Phi$ EOS only leads to an
enhancement in the disk mass \citep{Foucart2012}, consistent with the
fact that different EOSs will affect mostly the amount of remnant disk
mass, but hardly the spin of the BH or the low-density part of the EOS
\citep{Timmes2000} probed in the accretion disk. Tidal disruption depends
on the mass ratio and spin of the BH \citep{ShibataTaniguchi2008} (see
also \citet{ShibataTaniguchilrr-2011-6} for a review). More precisely,
depending on the different effective spins of the BH (and hence also mass
ratios), tidal disruption can be enhanced in our set of models
\citep{Most2020e}. Indeed, we find that spin enhances the disk mass as
expected from previous studies \citep{Foucart2018b}, creating the most
extended and massive disk for the high-spin system (left panel), with a
mass $M^{\rm disk}_b = 0.17\, M_\odot$ (see Tab. \ref{tab:initial}). At
the same time, the zero-spin system \texttt{TNT.chit.0.00} only reaches
disk masses of $M^{\rm disk}_b = 0.01\, M_\odot$. Interestingly, the
intermediate case with effective spin $\tilde{\chi} = 0.15$ features the
highest rest-mass densities of all cases. The magnetic-field strength is
highest in the high-spin case, and lowest in the case of zero BH spin. A
more detailed discussion of the magnetic-field evolution will be given in
Sec. \ref{sec:mag}. The temperature, $T$, also increases monotonically
with spin, albeit in all cases the disks are rather cold $T\lesssim 10\,
\rm MeV$, and cooling over time.

\begin{figure*}
  \centering
  \includegraphics[width=0.95\textwidth, trim=0. 2.4cm 0. 0. , clip]{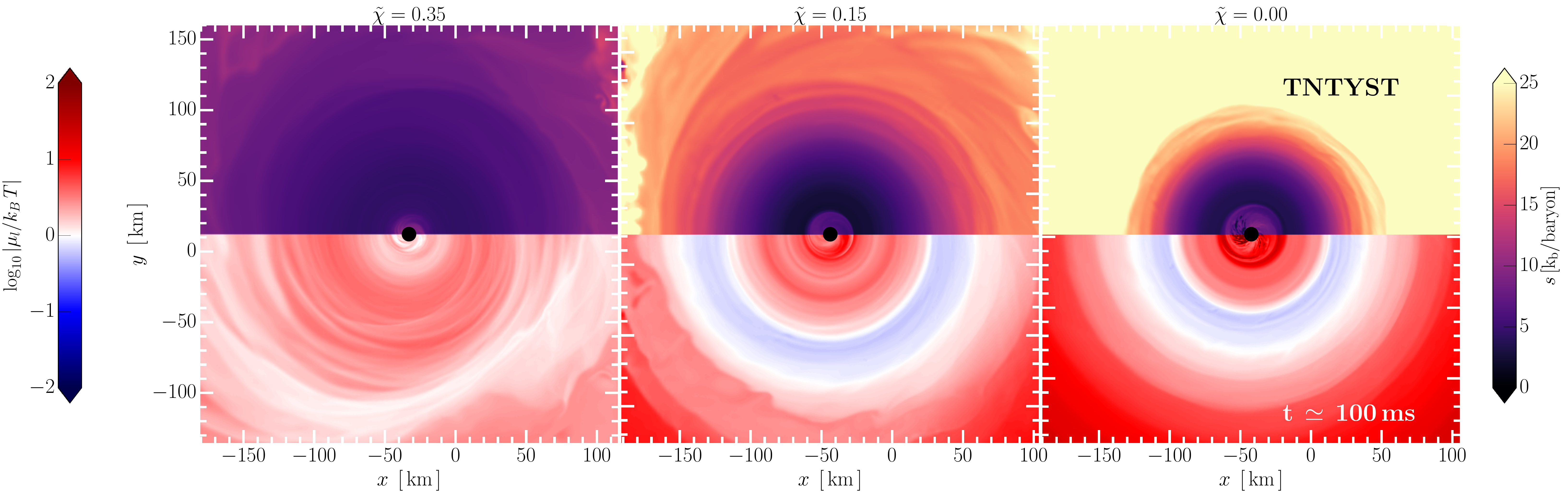}
  \includegraphics[width=0.95\textwidth, trim=0. 0. 0. 1.cm, clip]{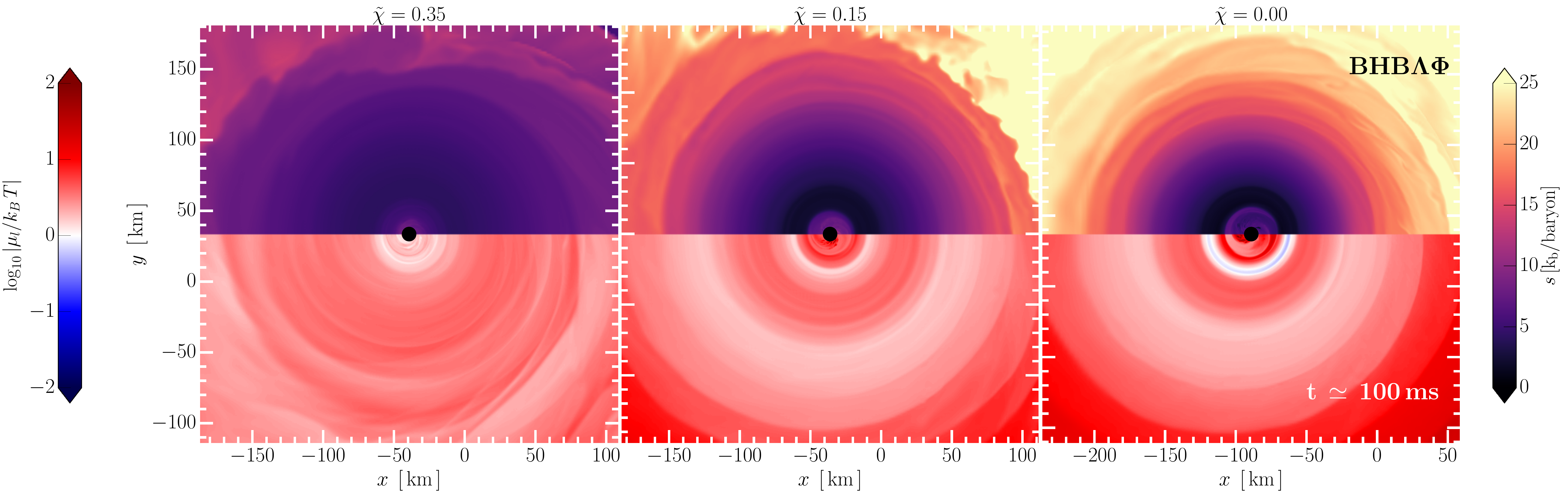}
  \caption{Equatorial view of the entropy $s$ per baryon and the lepton
    chemical potential $\mu_{\rm l}$, relative to the fluid temperature
    $T$. The top row shows models computed using the \texttt{TNTYST} EOS,
    the bottom row \texttt{BHB}$\Lambda\Phi$ models. All results are
    shown at $\simeq 100\, \rm ms$ after merger.}
  \label{fig:xy_muL}
\end{figure*}

\begin{figure}
  \centering
  \includegraphics[width=1.0\columnwidth]{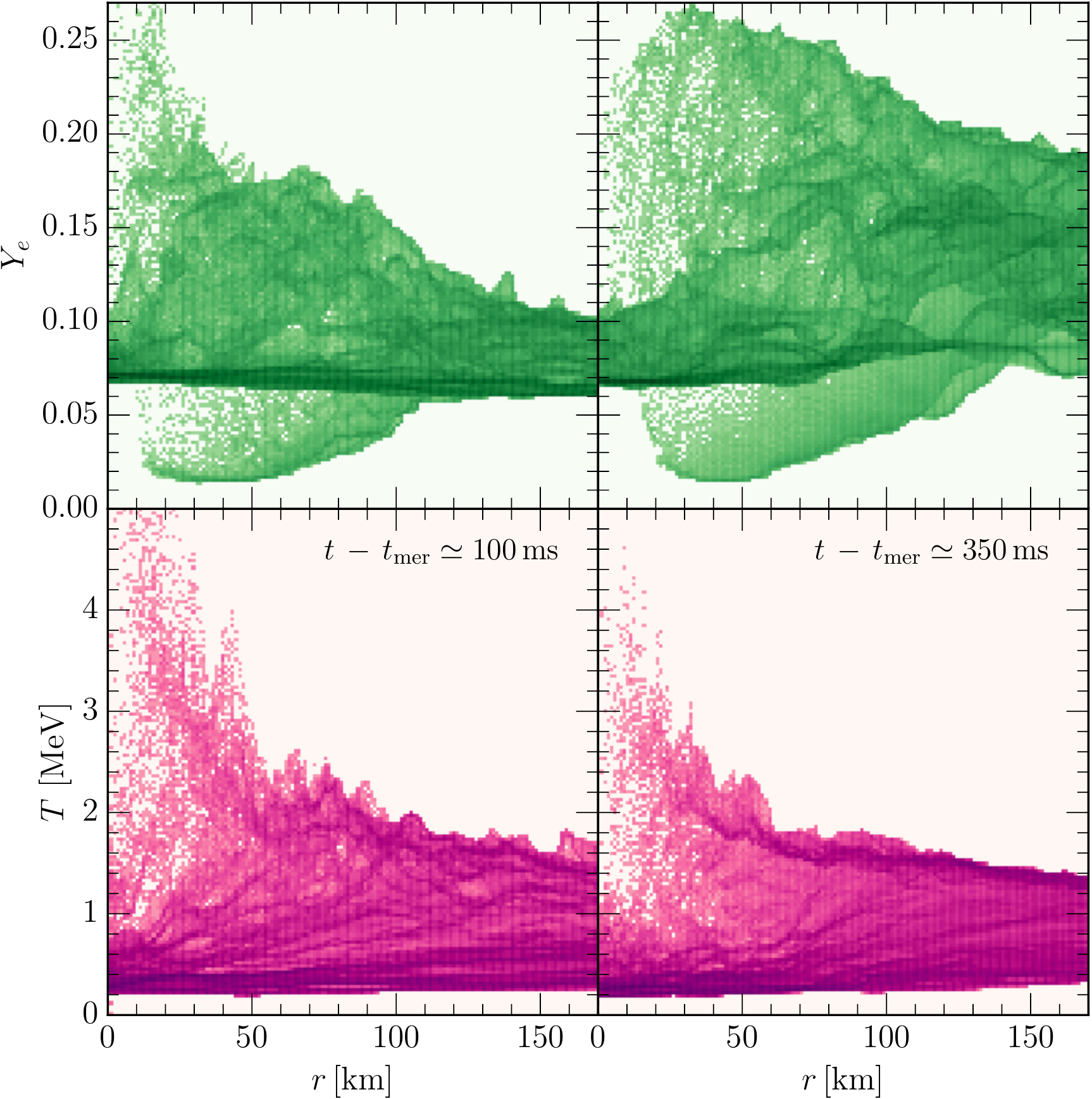}
  \caption{Electron fraction $Y_e$ (top panels) and temperature $T$
    (bottom panels) sampled as a function of the cylindrical radius $r$
    at two times after the merger. Shown are results for model
    \texttt{TNT.chit.0.35}.  }
  \label{fig:Y_ign}
\end{figure}

The continued emission of neutrinos leads to a rapid cooling of the disk,
which is shown in Fig. \ref{fig:hydro_prop_tntyst}. Different from the
collision of two NSs, where the compression at merger can produce very high
temperatures $\gg 10\, \rm MeV$ \citep{Perego2019,Endrizzi2020}, tidal
disruption is not able to significantly heat up the baryonic matter in
the simulation. In fact, when comparing the different evolutions of the
hottest fluid elements in the simulation domain (top panels of
Fig. \ref{fig:hydro_prop_tntyst}), we find that, at most, temperatures of
$15\, \rm MeV$ are reached in the case of \texttt{TNT.chit.0.35} and
about $10\, \rm MeV$ in the lower spin cases. Neutrino emission then
leads to a rapid cool-down of the disk to below $10\, \rm MeV$ on a
timescale of about $\sim 50\, \rm ms$. The magnitude of neutrino
luminosity from the disk is also less for the low-spin models (bottom
panels of Fig. \ref{fig:hydro_prop_tntyst}), consistent with the disks
being colder. In all cases the initial neutrino luminosity is at least
$10^{50}\, {\rm erg\,s^{-1}}$.

We can further see in Fig.~\ref{fig:xz_tntyst} that the ``funnel'' region
is mildly polluted with neutron-rich matter, except in the case of
low-spin \texttt{TNTYST} models, which feature lower densities at higher
electron fraction. This likely indicates that this matter is closer to
beta-equilibrium which corresponds to more symmetric matter at those
densities.

\begin{figure*}
  \centering
  \includegraphics[width=\textwidth]{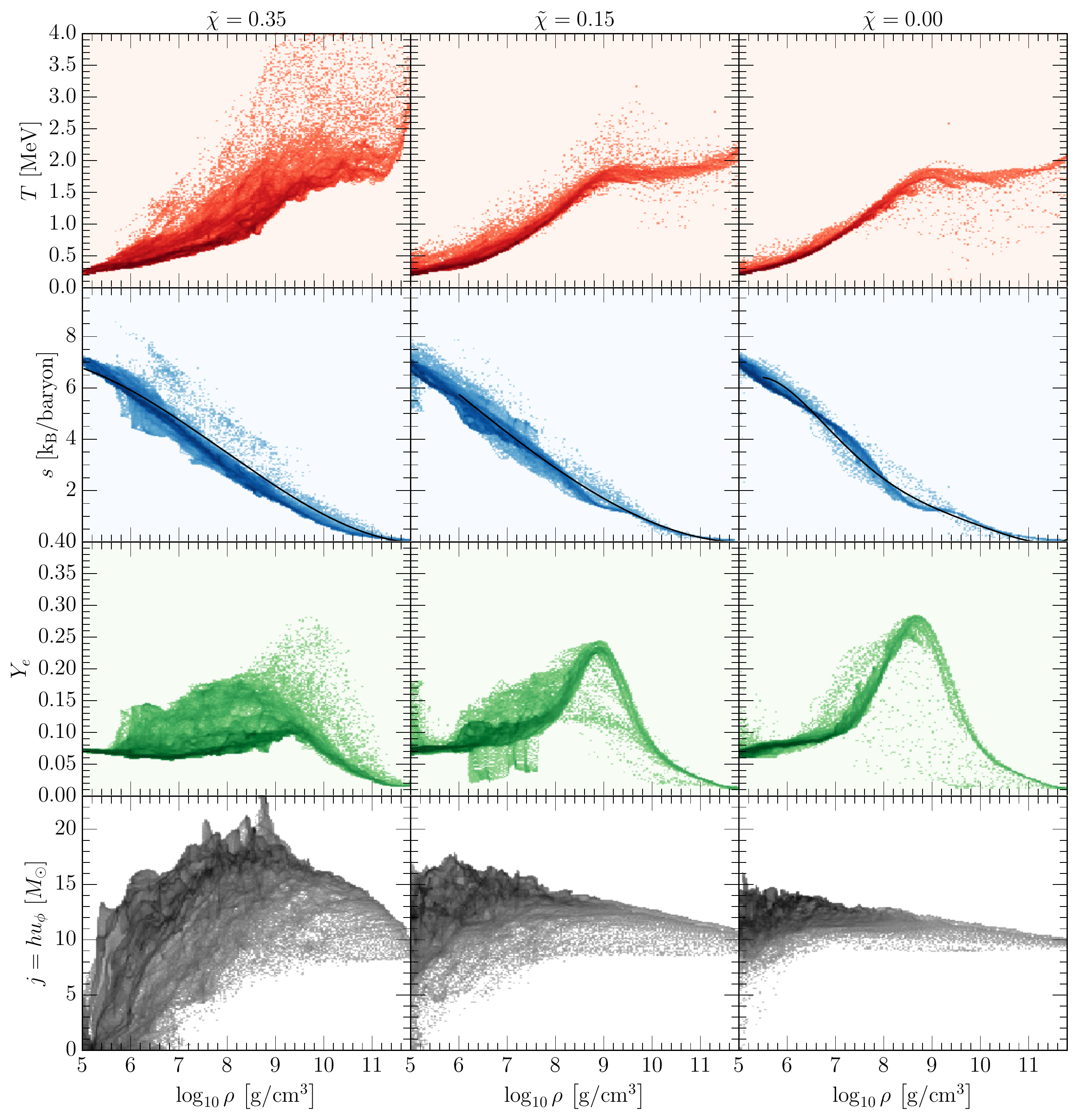}
  \caption{Properties of the accretion disks formed in BH -- NS merger
    simulations using the \texttt{TNTYST} EOS for different effective
    spins $\tilde{\chi}$ at the same time as in Fig. \ref{fig:xz_tntyst}
    ($t-t_{\rm mer} \simeq 100\, \rm ms$). Shown are the temperature $T$,
    specific entropy per baryon $s$, electron fraction $Y_e$ and specific
    angular momentum $j := h u_\phi$ sampled based on their local
    rest-mass density $\rho$.}
  \label{fig:Dp_TNT}
\end{figure*}

\subsubsection{Nuclear composition and weak-interactions}

When looking at the nuclear composition of the disks, in terms of the
electron fraction $Y_e$, we can see that the disks become more neutron
rich with increasing effective spin. Furthermore, the disks begin to
rapidly protonize in the outer layers, since these are transparent to
neutrino emission. This can easily be understood when considering that
the disks, initially formed from almost pure neutron matter in the tidal
disruption process, have a composition that is far from beta-equilibrium
under the post-merger conditions. Hence, beta-decay will lead to an
increase in the proton--neutron ratio in the disk, that is accompanied by
an emission of neutrinos. Quantitatively, this leads to an increase in
the electron fraction which easily reaches $Y_e \simeq 0.2-0.3$. While
this is most pronounced for the medium and zero spin cases (right and
middle columns in Fig. \ref{fig:xz_tntyst}), we expect that on larger
time and lengthscales the disk in the highest-spin case (left column)
will exhibit a similar behaviour.

This effect has been closely investigated by \citet{De2020}, who found
that starting with constant specific angular momentum and specific
entropy disk equilibria, low-mass disks protonize more quickly and that
weak interactions \textit{switch-off} in these disks. Different from
these idealized accretion disks, we find that the protonization is not
highest in the center around the BH but largely affects a ring of
low-density material. In Fig. \ref{fig:xy_muL} the spatial distribution
of the specific entropy $s$ and the lepton chemical potential $\mu_{\rm
  l}$ is reported, for which the latter tends to zero in
beta-equilibrium\footnote{We recall that for the conserved quantum
numbers charge ${Q}$, baryon number ${B}$, and lepton number ${l}$ [\eg a
  neutron is a baryon (${B}=1$, ${l}=0$) with zero electric charge
  (${Q}=0$)], the chemical potential will split as
\begin{align*}
  \mu = B \mu_{\rm B} + Q \mu_{\rm Q} + l \mu_{\rm l}\,,
\end{align*}
where $\mu_{\rm B, Q, l}$ are the associated chemical
potentials. Thus, for beta-equilibrium,
\begin{align*}
\mu_{\rm p} + \mu_{\rm e} - \mu_{\rm n} = \mu_{\rm l} \rightarrow 0\,,
\end{align*}
using the neutron chemical potential, $\mu_{\rm n} = \mu_{\rm B}$, the
proton chemical potential, $\mu_{\rm p} = \mu_{\rm B} + \mu_{\rm Q}$, and
the electron chemical potential $\mu_{\rm e} = \mu_{\rm l} - \mu_{\rm
  Q}$.}. Indeed, this proton-rich ring quickly beta-equilibrates (blue
regions), as $\mu_{\rm l}/k_B T \ll 1$.  Interestingly, we find that in
the case of the $\rm BHB\Lambda\Phi$ models (bottom row) an inner ring of
matter close to beta-equilibrium develops, whereas the outer ring is less
equilibrated than in the \texttt{TNTYST} cases. This inner ring is
seemingly absent for models with the \texttt{TNTYST} EOS (top
row). Moreover, we find that our realistic remnant accretion disks
feature density-dependent variations of the specific entropy, $s< 10$.

To complete our discussion on weak interactions in the disk, we briefly
point out the presence of disk self-regulation
\citep{Chen2007,Siegel2017}, following the discussion in
\citet{Siegel2018}. While Fig. \ref{fig:xy_muL} would imply that large
parts of the disk are out of beta-equilibrium and would have to
protonize, self-regulation at the inner edge of the disk, will lead to
neutronization that maintains $Y_e \simeq 0.1$. This is because in this
neutrino transparent regime copious $e^\pm$-pair production will
effectively modify the beta-equilibrium condition
\citep{Beloborodov2003}. The balance between turbulent heating (induced
by the magneto-rotational instability, see Sec. \ref{sec:mag}) and
neutrino cooling will then establish this reservoir of neutron-rich
matter in the inner edge of the disk.

Although this study primarily focusses on obtaining initial conditions
for disks after tidal disruption, our long-term simulation
(\texttt{TNT.chit.0.35}), was run long enough to exhibit this behaviour.
Indeed, as shown in the top panels of Fig. \ref{fig:Y_ign}, the disk is
initially (\ie at $t-t_{\rm mer}=100\, \rm ms$) very neutron rich with
$Y_e<0.15$, since it was formed from the neutron-rich matter of the tidal
disrupted NS. However, at the final time of our simulation (\ie at
$t-t_{\rm mer}=350\, \rm ms$), large parts of the disk have already begun
to protonize, reaching $Y_e\simeq 0.2$ at larger distances $r > 80\,\rm
km$ from the BH, where densities are larger and the cooling less
effective. We can also see this by looking at the temperature profiles in
the bottom panels of Fig. \ref{fig:Y_ign}. More specifically, at later
times while the main part of the disk cools only slowly, close to the BH
and at distances $r < 40\,\rm km$, where the densities are smaller and
the cooling is more effective, most of the disk matter is even more
neutron rich than initially, \ie having $Y_e < 0.1$. This is consistent
with the onset of self-regulation and broadly agrees with the findings
presented in \citet{Siegel2018}.

\begin{figure*}
  \centering
  \includegraphics[width=0.9\textwidth]{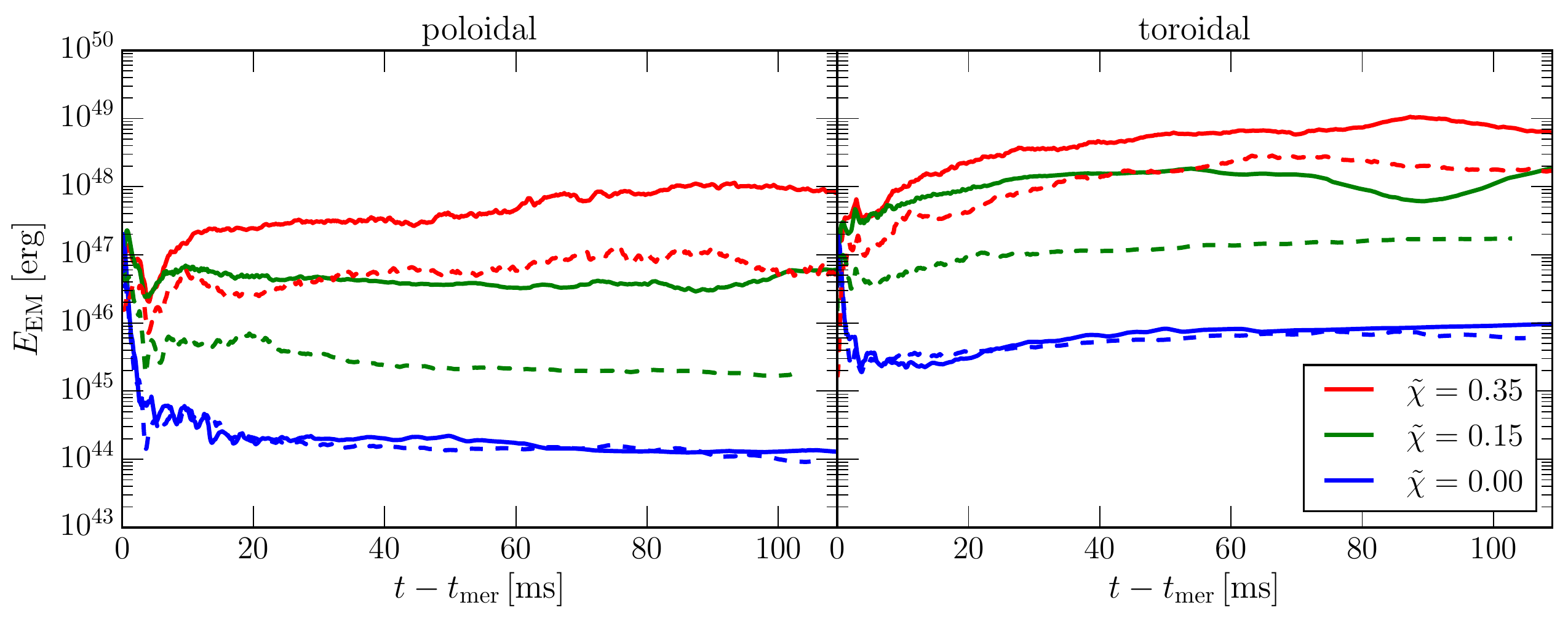}
  \caption{Evolution of the magnetic energy, $E_{\rm EM}$, after the
    merger at time $t_{\rm mer}$. Shown are the total poloidal (left) and
    toroidal component (right). The colors refer to the different models
    in terms of their effective spins, $\tilde{\chi}$. Solid lines
    represent simulations with the \texttt{TNTYST} and dashed lines with
    the \texttt{BHB}$\Lambda\Phi$ EOSs. }
  \label{fig:em_tntyst}
\end{figure*}

\subsection{Bulk hydrodynamic properties}
\label{sec:hydro_prop}

We now focus on the hydrodynamic properties of the accretion disk. When
constructed as equilibrium configurations in axisymmetry
\citep{Fishbone76}, accretion disks are usually described in terms of the
specific angular momenta $j$, specific entropies $s$, disk masses
$M_{\rm disk}$, and their electron fractions $Y_e$. Typically, those are
assumed to be constants in most previous studies
(e.g. \citet{Siegel2017,Fernandez2018}). Having a set of fully consistent GRMHD
simulations of accretion disks formed by tidal disruption in the merger
of BH--NS systems, does in fact allow us to compute realistic
distributions of the above fluid quantities as functions of the rest-mass
density $\rho$. These are shown in Fig. \ref{fig:Dp_TNT} for simulations
using the \texttt{TNTYST} EOS. Starting from the top, we first show the
temperature distributions that, in all of the three cases, exhibit a flat
plateau around $T\simeq 2\, \rm MeV$, which changes to a fall-off for
rest-mass densities $\rho <10^9 \, \rm g \, cm^{-3}$. The entropy
profiles (second row from the top) are almost linear in all cases,
starting from specific entropies $s<1$ at the highest densities and
extending outwards until $s\approx 7-8$ for the lowest densities in the
disks, $\rho \simeq 10^5\, \rm g \, cm^{-3}$. Interestingly, low-mass
disks, \ie obtained for $\tilde{\chi}= \left[ 0.00\,, 0.15 \right]$, seem
to feature a small plateau around $\rho \simeq 10^9\, \rm g\, cm^{-3}$,
corresponding roughly to the transition to the nuclear statistical
equilibrium EOS used at low densities. Overall, it is possible to provide
an effective fit for the specific entropy in terms of the function
\begin{align}
  s = \sum_{k=0}^5 c_k \left(\log_{10}\left[\rho/ \left( \rm g\,cm^{-3} \right) \right]\right)^k\,.
  \label{eqn:sfit}
\end{align}
where the numerical values for the coefficients $c_k$ are provided in
Tab. \ref{tab:fit}.

\begin{table*}
  \centering
  \begin{tabular}{{|l|c|c|c|c|c|c|c|c|}}
    \hline
    \hline
    &
    $c_0$  &
    $c_1$  &
    $c_2$  &
    $c_3$  &
    $c_4$  &
    $c_5$  &
    $n_{\rm P}$  &
    $n_{\rm T}$  \\
    \hline
    \hline
  \texttt{TNT.chit.0.00} & $-2.87\times 10^{2}$ &$1.81\times 10^{2}$ &
  $-4.32\times 10^{1}$ & $5.01$ & $-2.85\times 10^{-1}$ & $6.41 \times
  10^{-3}$ & $0.63$ & $0.81$ \\
  \texttt{TNT.chit.0.15} & $1.32\times 10^{1}$ & $-5.15\times 10^{-1}$&
  $-1.92\times 10^{-1}$ & $1.19\times 10^{-2}$ & $-$ & $-$ & $0.60$ & $0.83$ \\
  \texttt{TNT.chit.0.35} & $1.27$ &$3.50$ &
  $-6.09\times 10^{-1}$ & $2.56\times 10^{-2}$ & $-$ & $-$ & $0.68$ &
  $0.81$ \\
       \hline                                                                             
  \texttt{BHB.chit.0.00} & $-2.21\times 10^{2}$ &$1.44\times 10^{2}$ &
  $-3.53\times 10^{1}$ & $4.17$ & $-2.42\times 10^{-1}$ & $5.53 \times
  10^{-3}$ & $0.67$ & $0.87$ \\
  \texttt{BHB.chit.0.15} & $2.21\times 10^{1}$ & $-3.16$& $6.72\times 10^{-2}$ &
  $3.55\times 10^{-3}$ & $-$ & $-$ & $0.77$ & $0.97$ \\
  \texttt{BHB.chit.0.35} & $8.39$ &$1.39$ &
  $-4.15\times 10^{-1}$ & $2.03\times 10^{-2}$ & $-$ & $-$ & $0.67$ & $0.87$ \\
       \hline                                                                                   \hline                                                                              \end{tabular}
  \caption{Numerical coefficients for the fits of the specific entropy
    $s$ and magnetic-field strengths $B_{\rm P,T}$. See
    Eqs. \eqref{eqn:sfit} and \eqref{eqn:Bfit} for details.}
    \label{tab:fit}
\end{table*}

Focussing on the electron fraction $Y_e$, we can continue our earlier
discussion around Fig. \ref{fig:xy_muL}, where a ring-like structure in
the disk approaching beta-equilibrium is observable. Looking at $Y_e
\left( \rho \right)$ (third row from the top in Fig.~\ref{fig:Dp_TNT}),
we can see that the bulk of the matter in the disk follows a tight
relation with the rest-mass density. The innermost, densest parts of the
disk, \ie $\rho > 10^{10}\,\rm g \, cm^{-3}$, are at very low electron
fractions, $Y_e < 0.05$, thus implying that most of the disk is made of
almost pure neutron matter. Moving outwards to lower densities and
approaching the beta-equilibrated ring (for $\tilde{\chi} = 0.00, 0.15$),
the electron fraction peaks at $Y_e = 0.25$ for densities around $\rho =
10^9 \rm \, g\, cm^{-3}$. Instead of continuing to increase as would be
predicted in beta-equilibrium\footnote{Recall that at low-densities
matter will be approximately symmetric and, therefore, $Y_e \simeq
0.5$.}, the electron fraction decreases again to below $Y_e<0.1$. For the
highest-spin case, the situation is slightly different. We can see that a
small fraction of the disk material exhibits a similar
(beta-equilibrated) $Y_e$-peak as in the other cases, but most of the
disk remains at electron fractions $Y_e<0.1$, indicating that the entire
disk remains neutron rich. This is consistent with the weak-interaction
``ignition threshold'' proposed by \citet{De2020}.

Concerning the specific angular momentum that governs the structure of
the disk (bottom row), we find that depending on the effective spin
$\tilde{\chi}$ and, hence, on the disk mass, the distributions look very
different. Although there are different definitions of the relativistic angular
momentum $j$ in use in the literature \citep{Kozlowski1978},  we
have found only minor differences between them when
extracted from our simulations. For simplicity, we therefore choose 
to adopt the one used for massive disks
$j:= h u_\phi$, where $h$ is the specific enthalpy and $u_\phi$ the
covariant azimuthal-component of the fluid four-velocity \citep{Kiuchi2011b}.
For the highest disk mass and spin (left panel), we find for large parts of
the disk,
$\rho > 10^{8}\, \rm g \, cm^{-3}$, that the specific angular momentum is
far from being constant, as customarily assumed in simplified models of
equilibrium tori. Rather, the specific angular momentum varies in the
range $ 10 < h u_\phi /M_\odot < 20$, and decreases further out at low
densities $<10^8\,{\rm g\,cm^{-3}}$ as required by dynamical
stability. Only for lower spins and disk masses, can the specific angular
momentum be considered closer to a constant, although also in this case
it ranges from $ 10 < h u_\phi/ M_\odot < 15$.

In comparison, Fig. \ref{fig:Dp_BHB} in Appendix \ref{sec:appendix_a},
shows the same distributions but for the mergers using the
\texttt{BHB}$\Lambda\Phi$ EOS. While the temperature, entropy and
electron fraction distributions look overall very similar, the angular
momentum distributions are different. For high and medium spin (left and
middle panel, bottom row), the results do look comparable, but the
zero-spin case features a split of the specific angular momentum in two
semi-constant branches.

\begin{figure*}
  \centering
  \includegraphics[width=0.9\textwidth]{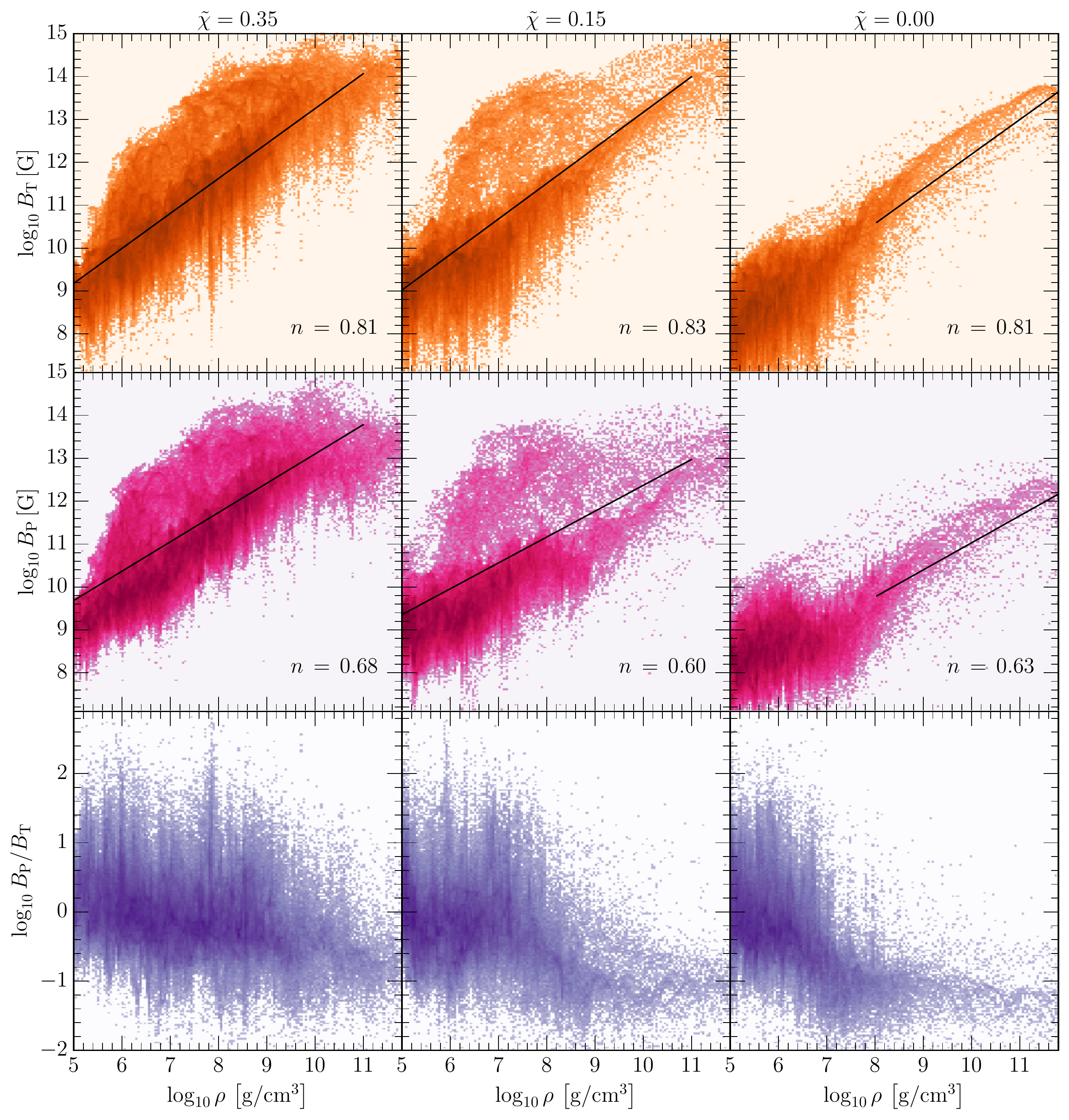}
  \caption{Density-dependent distributions of the poloidal, $B_{\rm P}$,
    and toroidal, $B_{\rm T}$, components of the magnetic field at the
    same time as in Fig. \ref{fig:xz_tntyst} ($t-t_{\rm mer} \simeq 100\,
    \rm ms$). The simulations shown use the \texttt{TNTYST} EOS and $n$
    is defined in Eq. \eqref{eqn:Bfit}.  }
  \label{fig:Bp_TNT}
\end{figure*}

\begin{figure*}
  \centering
  \includegraphics[width=\textwidth]{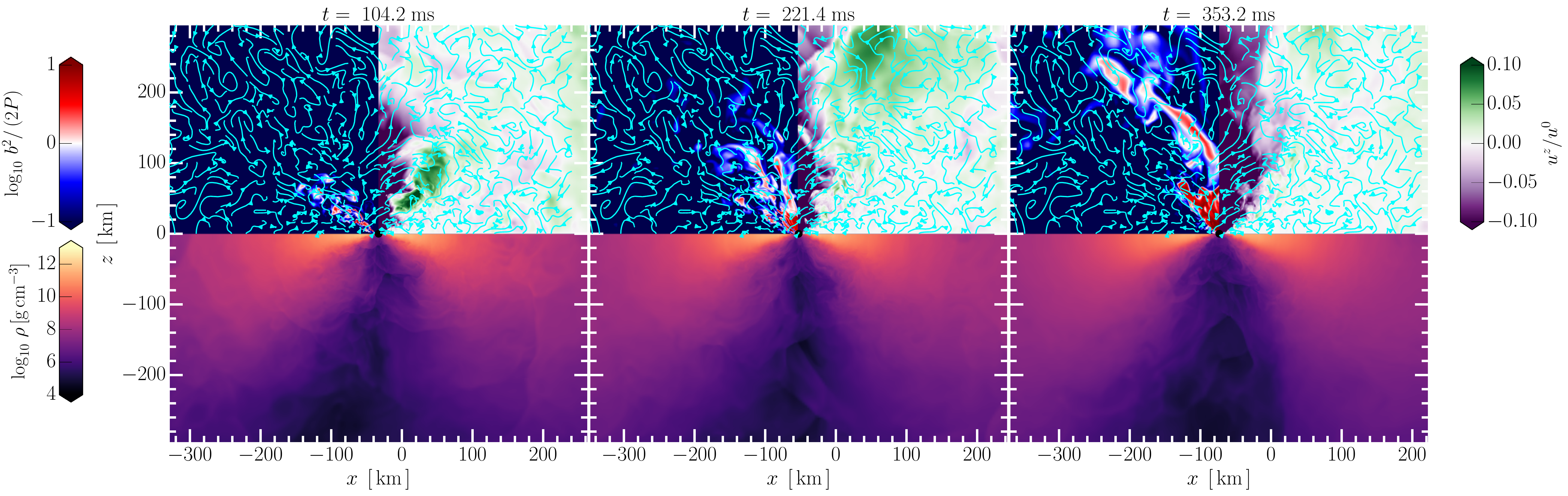}
  \caption{Long-term evolution of the \texttt{TNTYST.chi.035} system. The
    top half shows the vertical flow velocity $u^z/u^0$ and the inverse
    plasma parameter $\beta := b^2/\left(2 p\right)$, computed in terms
    of the comoving magnetic energy density $b^2$ and of the fluid
    pressure $p$. The lower half of the panel shows the rest-mass density
    $\rho$. The magnetization of the funnel is steadily growing over time
    (red regions), while the matter inside the funnel is increasingly
    inflowing (violet regions), indicating potential funnel clearing on
    longer timescales.}
  \label{fig:LTb2}
\end{figure*}

\begin{figure}
  \centering
  \includegraphics[width=0.45\textwidth]{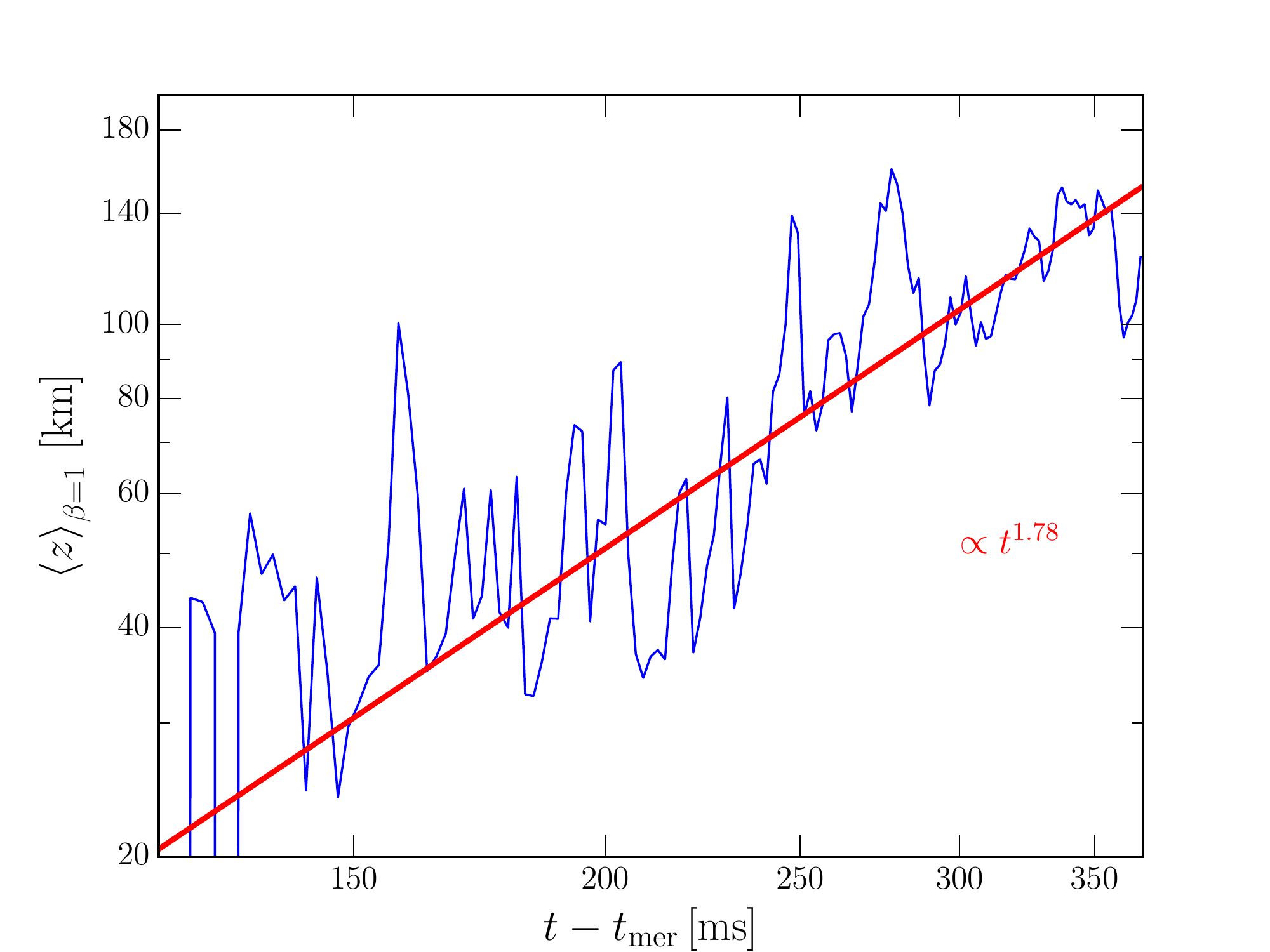}
  \caption{Characteristic scale-height of the magnetized funnel region, where
    $\langle z \rangle_{\beta=1}$ refers to the average vertical height
    corresponding to the plasma parameter $\beta = b^2/\left(2 p\right) =
    1$. The time is measured relative to the time of merger $t_{\rm mer}$.}
  \label{fig:b2pZ}
\end{figure}

\subsection{Magnetic-field evolution}
\label{sec:mag}

While the disk is cooling, the dynamics of the plasma and magnetic
instabilities lead to an increase in the magnetic field strength, as shown in
Fig. \ref{fig:em_tntyst}, which displays the evolution of the toroidal
and poloidal magnetic energy. Note that after the merger, the disruption
of the NS leads to a shearing of the magnetic-field lines in its
interior, which causes a sudden amplification of the magnetic field
\citep{Rezzolla:2011, Etienne2012}. Because of the geometry of the
disruption event, the amplification will mainly affect the toroidal
component of the field. The predominantly poloidal parts of the magnetic
fields in the center of the disrupted star experience a smaller
amplification and are mostly accreted by the BH, as can be seen by the
sudden drop in poloidal energy directly after merger (left panel). In the
case of the zero-spin binary \texttt{TNT.chit.0.00}, almost the entire NS
is accreted and only the weakly magnetized matter in the outermost parts
of the original NS remain to form the disk. This is accompanied with a
very sharp drop in both components of the magnetic energy. After the
merger the magnetorotational instability (MRI) \citep{Velikhov1959,
  Chandrasekhar1960, Balbus1991} will begin to drive an early
amplification of the magnetic field, which can be seen by an increase in
both poloidal and toroidal field components. A key feature to accurately
capture this amplification is with the use of a fourth-order accurate
numerical scheme \citep{Most2019b}, which helps to resolve instabilities,
such as the MRI, even at lower resolutions in the outer parts of the
disk. The use of such schemes can, however, not alleviate the need for
resolving the lengthscales associated with physical processes and
instabilities. Indeed, we find that the lowest mass disk, case
\texttt{TNT.chit.0.00}, leads to a disk of roughly half the size of the
other cases, pointing to the need for higher resolutions than we
currently employ. The absence of poloidal field growth for this case
confirms that the MRI wavelength is not fully resolved in large parts of
the disk and that the MRI is likely not active in this simulation. Given
the low mass of the disk and the low observational prospects, we choose
to not repeat this calculation at higher resolution.

Comparing the evolution of the magnetic energy for different BH spins, we
find that the high-spin cases feature larger magnetic energies, with the
models with the \texttt{TNTYST} EOS (solid curves), having up to an order
of magnitude higher electromagnetic energies than the
\texttt{BHB}$\Lambda\Phi$ models (dashed curves). This already points to
a mild EOS dependence on the subsequent disk evolution. Moreover, we can
already anticipate that the magnetic-field geometry is largely toroidal
by comparing the left and right panels in Fig. \ref{fig:em_tntyst}. Since
the initial structure of the field geometry might critically affect the
timescales of the late-time evolution of the disk \citep{Christie2019b},
we next present a detailed analysis of the magnetic-field topology.
\begin{figure*}
  \centering
  \includegraphics[width=0.75\textwidth]{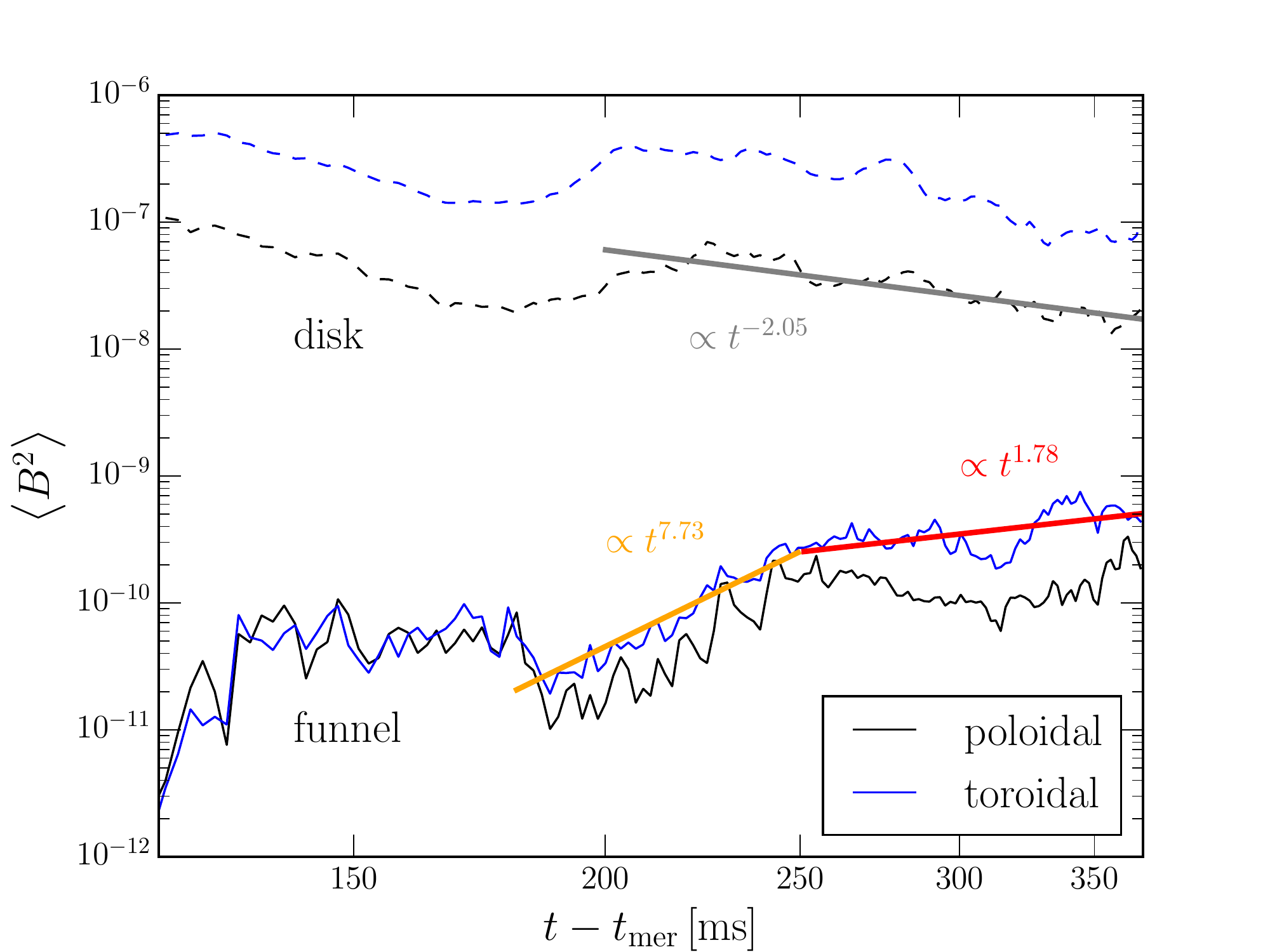}
  \caption{Evolution of the mean magnetic energy density, $\langle
    B^2\rangle$, in terms
  of poloidal (black) and toroidal (blue) components. The evolution is
  shown separately for the disk (dashed lines) and for the funnel region
  (solid lines). The time is measured relative to the time of merger
  $t_{\rm mer}$.}
  \label{fig:topol}
\end{figure*}

\subsection{Magnetic-field topology}

We can further investigate the structure and properties of the magnetic
field by looking at density distributions of the poloidal $B_{\rm P}$ and
toroidal $B_{\rm T}$ magnetic-field components, as shown in
Fig. \ref{fig:Bp_TNT} for the simulations using the \texttt{TNTYST}
EOS. Starting from the irrotational case, $\tilde{\chi}=0.00$ (left
column), we can see that, overall, the bulk of the poloidal and toroidal
fields follows a power-law dependence with the rest-mass density
\begin{align}
  \log_{10} \left[B_{\rm P,T}/{\rm G}\right] = n_{\rm P,T} \log_{10} \left[\rho /
\left({\rm g\, cm^{-3}}\right)\right] + \rm const.\,,
  \label{eqn:Bfit}
\end{align}
where the values of the two coefficients $n_{\rm P,T}$ are reported in
Tab. \ref{tab:fit}.

Note that the toroidal magnetic field peaks at $10^{14}\, G$. The
poloidal to toroidal ratio is $B_{\rm P}/B_{\rm T}\simeq 0.1$, and
increases locally to $B_{\rm P}/B_{\rm T} \lesssim 10$ for low rest-mass
densities $\rho < 10^7\, \rm g\, cm^{-3}$. This is mainly due to the
appearance of the MRI, which causes a sustained replenishing of the
poloidal field \citep{Sadowski2015}. Conversely, this also confirms our
initial conclusion that the MRI is likely not resolved in the least
massive and, hence, smallest of the disks. While this clearly indicates
that any subsequent evolution of this disk at current resolution is not
feasible, it also allows us to draw an important conclusion about the
correct initial conditions for the magnetic field. Namely, that the
profiles seen in the third column of Fig. \ref{fig:Bp_TNT} should be
indicative for realistic initial magnetic-field topologies in disks
formed directly in BH-NS mergers. Since the subsequent magnetic-field
evolution is very modest, these distributions represent the initial
magnetic-field configuration in the disk as soon is it equilibrates after
merger. More importantly, these distributions are rather different from
those normally employed in simulations starting from axisymmetric tori in
equilibrium, which often even ignore the presence of a toroidal
component.

This qualitative behaviour described before is the same for all
disks. However, the larger poloidal-toroidal ratio is indicative of the
fact that the MRI is active in increasingly larger parts -- and at higher
densities -- of the disks. For the high-spin case, $\tilde{\chi}=0.35$,
we can clearly see that the MRI is active throughout the disk and that
both poloidal and toroidal field grow beyond the simple power-law
behaviour outlined before. Looking at the corresponding
Fig. \ref{fig:Bp_BHB} in Appendix \ref{sec:appendix_a}, it is possible to
deduce that this behaviour holds also for a different EOS, and that
the overall magnitude of the magnetic field remains insensitive to the
initial choice of EOS for the inspiraling NS. Nonetheless, the slope
coefficients $n_{\rm P,T}$ are slightly larger when compared to the
simulations with the \texttt{TNTYST} EOS. Although this change is rather
minor, it does hint that the slightly altered distribution of the
magnetic field inside the initial NS is partially imprinted onto the
disk, as it would be natural to expect.

\subsection{Prospects of jet launching}

Having discussed the composition structure of the disk and the magnetic
field topology present in the accretion disks formed in (low-mass) BH--NS
mergers, we now turn to the prospect of launching a jet from these
systems. While this process has been thoroughly studied for accretion
disks found in supermassive BH accretion \citep{Abramowicz2011,
  Porth2019, Davis2020}, only a few attempts have been made in the
context of NS mergers, both in ideal \citep{Rezzolla:2011,
  Paschalidis2014, Kiuchi2015, Kawamura2016, Ruiz2018} and resistive MHD
\cite{Dionysopoulou2015, Qian2018}. Unless an external field was
initially seeded by means of a force-free like magnetosphere
\citep{Paschalidis2014,Ruiz2020b}, most simulations have
only observed the formation of
a helical magnetic-field structure in the funnel region
\citep{Kawamura2016}, however, not an actual (relativistic) outflow. In
most cases, strong baryon pollution from the disk created large ram
pressures preventing the funnel from clearing and attaining a
magnetically dominated, force-free state \cite{Kiuchi2014,
  Kiuchi2015}. Since most of these simulations were run with low
resolutions and for short timescales $\lesssim 100\, \rm ms$, it
currently remains an open problem to understand under what conditions
the magnetization in the funnel would grow over longer timescales. Recent simulations of
accretion disks with toroidal magnetic fields indeed indicated that the
timescale for magnetic-field growth is significantly longer than for
initial poloidal geometries typically used to study jet launching
\citep{Liska2018b,Christie2019b}. While these studies benefit from having
higher resolutions close to the BH horizon, due to the use of better
suited spherical coordinate systems, the use of a fully fourth-order
numerical scheme allows us to accurately capture magnetic instabilities
in the disk with fewer grid points than needed for second-order codes
\citep{Most2019b}. Therefore, we evolve one of the systems
\texttt{TNTYST.chi.035} until $\sim 350\, \rm ms$ in order to gauge its
prospect for jet launching. Since the BH retains a net linear momentum
after merger, we continue to solve the Einstein equations alongside those
of GRMHD.

In Fig. \ref{fig:LTb2} we report the evolution of the funnel region in
terms of plasma parameter $\beta^{-1} := b^2/\left(2 p\right)$ -- where
$b^2$ is the comoving magnetic energy density and $p$ the fluid pressure
-- and the vertical flow velocity $u^z/u^0$. Note that while the funnel
starts out being only weakly magnetized (left panel), the magnetization
grows steadily over time with increasingly larger portions exhibiting
$\beta^{-1} >1$ (middle and right panels) in the polar regions above the
BH (funnel). The projected magnetic-field lines indicate a twistor shape
field geometry as observed previously by \citet{Kawamura2016}. At the
same time, the funnel regions begins to evacuate on essentially the same
timescale. A steadily increasing inflow, indicated by the dark violet
regions in Fig. \ref{fig:LTb2}, strongly hints to an eventual clearing of
the funnel, although large patches are still strongly polluted by disk
inflows (lower half of the panels). In order to better quantify the
growing magnetization of the funnel, we introduce a characteristic
scale-height $\langle z \rangle_{\beta=1}$, which corresponds to the
average $z$-coordinate with the plasma parameter $\beta =1$. This
parameter may be interpreted as a proxy for how the magnetization of the
funnel varies in the vertical direction. We caution that a fully
force-free funnel would require $\beta \ll 1$, although such values will
in practice not be present until shortly before jet launching. We show
the corresponding evolution of the scale-height in
Fig. \ref{fig:b2pZ}. Although there are inherent fluctuations over time,
on average we find that the magnetization scale-height is constantly
growing with a power law $\langle z \rangle_{\beta=1} \propto t^\alpha$,
with $\alpha = 1.78$. At the end of our simulation, this scale-height
extends to about $140\, \rm km$. If this growth was sustained at the same
rate, the scale-height would double around $\simeq 500\, \rm ms$. Since
the rest-mass densities decrease at larger distances from the BH, it
might be possible for this growth to accelerate at sufficient distance
from the BH, due to the faster decrease in pressure in those regions.

In order to better understand this behaviour, we have performed a
detailed analysis of the magnetic-field topology in the funnel and the
disk. To be more precise, we have computed mean magnetic-field energy
densities $\langle B^2\rangle$ for the toroidal and poloidal components
in the funnel and in the disk. We define the funnel region simply in
terms of density and angular cut-offs. More specifically, we define the
funnel to be within $30^\circ$ from the polar axis and contain densities of
at most $10^9\, \rm g\, cm^{-3}$. The resulting
evolution is shown in Fig. \ref{fig:topol}. We can see that initially the
magnetic-field strength is not growing in the funnel region. Only after
$\simeq 200\, \rm ms$, when the funnel has begun to clear we do see a
rapid growth in magnetic energy following a power law with exponent
$\alpha \approx 7.73$, which continues for around $50\, \rm
ms$. Afterwards the growth does not stop, but continues at a lower
rate. Surprisingly, it turns out that the growth is comparable to the
growth of the magnetization scale-height, having an exponent $\alpha
\approx 1.78$, consistent with the scaling reported in Fig. \ref{fig:b2pZ}.
At the same time the magnetic energy density in the disk is decreasing,
likely because of accretion regions with strong magnetic fields in the
accretion disks, which are closest to the ISCO, see Fig.
\ref{fig:xz_tntyst}.

\begin{figure}
  \centering
  \includegraphics[width=0.45\textwidth]{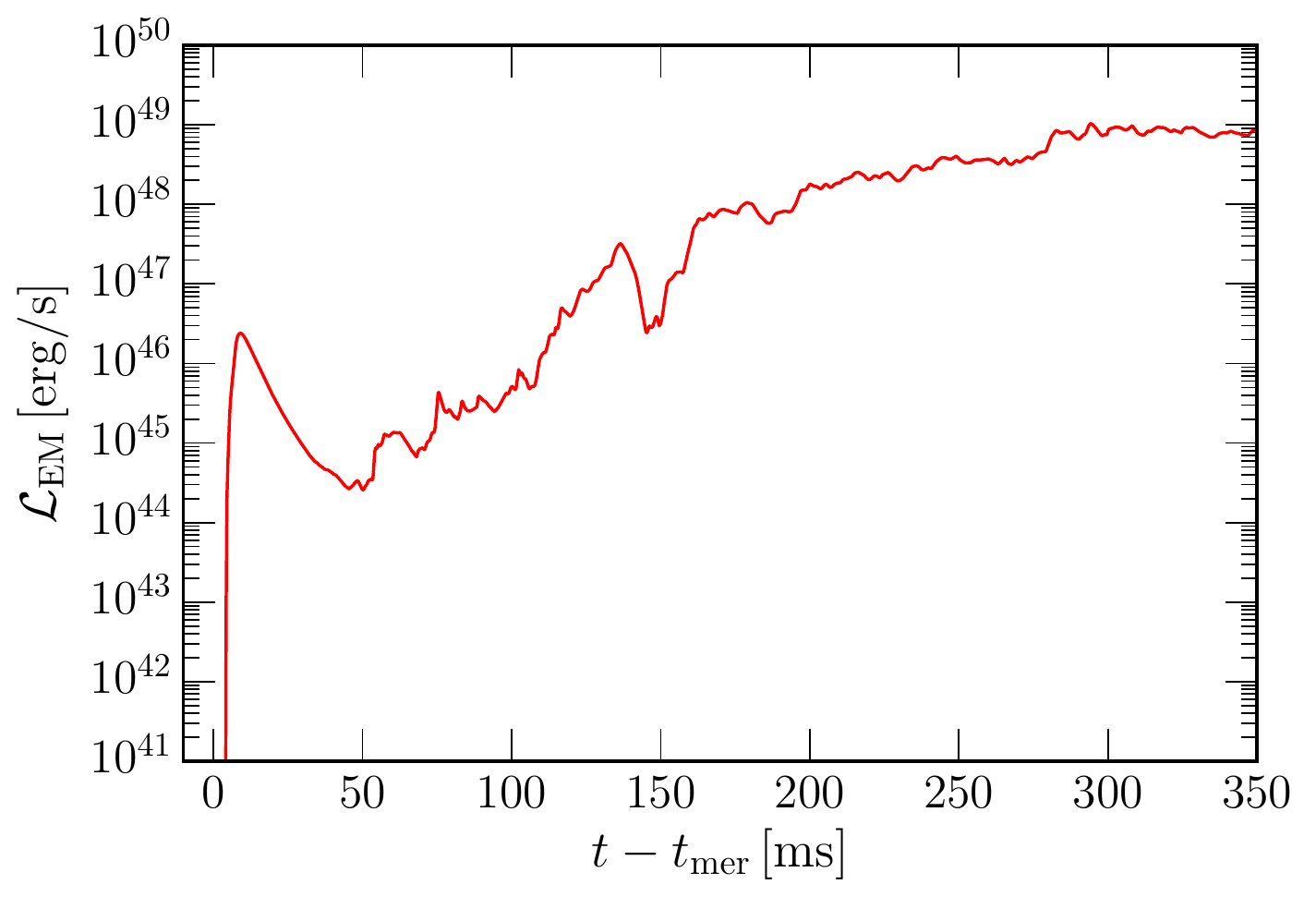}
  \caption{Electromagnetic luminosity $\mathcal{L}_{\rm EM}$ in terms of
    the Poynting flux for the long-term evolution of model
    \texttt{TNTYST.chi.035}. The time is measured relative to the time of
    merger $t_{\rm mer}$.}
  \label{fig:poynt}
\end{figure}

Finally, we also consider the electromagnetic energy outflowing from the
system. This is quantified in terms of the Poynting flux
\begin{align}
  \mathcal{L}_{\rm EM} = \oint T^t_{i\,,{\ \rm EM}} {d}S^i\,,
  \label{eqn:LEM}
\end{align}
extracted on a spherical surface at around $500\, \rm km$ from the BH.
Here $T^{\mu\nu}_{\rm EM}$ is the electromagnetic part of the
stress-energy tensor \citep{Baumgarte2003}. The resulting luminosity is
shown in Fig. \ref{fig:poynt}. Starting at $10^{45}{\rm erg\,s^{-1}}$ at
merger, the Poynting flux steadily increases and saturates at
$10^{49}{\rm erg\,s^{-1}}$ after about $300\, \rm ms$.

\section{Conclusions}

We have presented the results of a series of simulations in full general
relativity leading to the formation of accretion disks in the aftermath
of the merger of BH-NS systems in the near equal-mass regime parametrized
by the maximum mass of a nonrotating NS. Including strong magnetic
fields, weak interactions and realistic EOS, the disks have been evolved
for $\simeq 100\, \rm ms$ after merger until a quasi-stationary disk
equilibrium has been established. We have then provided a detailed
comparison of the disk properties for three different disk masses and BH
spins. In particular, we have found that the disks, while remaining very
cold, $T \lesssim 2\, \rm MeV$, have specific-entropy distributions that
follow power-laws in terms of the rest-mass density, $s \propto
\rho^{-n}$. Similarly, we were able to confirm that light disks more quickly
beta-equilibrate, while more massive disks remain neutron rich
\citep{De2020}. Importantly, massive disks have specific angular-momentum
distributions that are far from being constant, as instead customarily
assumed in simplified simulations starting from axisymmetric tori in
equilibrium, and vary in a rather wide range. At the same time, the
specific angular momentum distributions of the lighter disks have
smaller ranges of variation and may be roughly approximated as constant.

We have also found that the use of different EOSs with significantly
different compactnesses leads to changes in the disk mass, while the
other properties of the disk seem to be largely unaffected. Having
performed simulations with strong magnetic fields in the interior of the
NS companion, has allowed us to examine the magnetic-field structure
present in the post-merger disk when a quasi-stationarity solution is
reached. While previous studies have usually superimposed poloidal
magnetic fields \citep{Nouri2018} or -- more realistically -- toroidal
fields \citep{Christie2019b} on post-merger disks, we were able to find
density-dependent scaling laws for realistic magnetic-field
configurations, \ie $B\propto \rho^m$. Consistent with previous
simulations \citep{Giacomazzo:2010, Rezzolla:2011, Kiuchi2015} and
studies of binary NS mergers \citep{Kiuchi2014,Kawamura2016}, we find
that the field topology is initially strongly toroidal, albeit the onset
of the MRI leads to an eventual increase and amplification of the
poloidal field.

Although we believe that the results presented in this work will be
crucial for future modeling of post-merger accretion disks, some remarks
are in order. Owing to the high computational cost of performing BH-NS
mergers with accurate descriptions of the microphysics and magnetic-field
evolution, we have only investigated the near equal-mass regime. While
this might be most indicative for low-mass BH-NS and high-mass NS-NS
mergers (see \citet{Most2020d} for a comparison), realistic BH-NS mergers
are expected to happen for mass ratios $q<1/4$
\citep{Kruckow2018}. Although disk formation for these systems is largely
governed by the spin of the BH \citep{Foucart2012,Foucart2018b}, it
remains to be confirmed if the magnetic-field topology and entropy
profiles of the disk found here continue to hold also for these
systems. Since the primary formation mechanism, \ie tidal disruption,
operates similarly in all cases, we conjecture that most likely our
results should be transferable and therefore also hold qualitatively.

Finally, we have investigated the prospects of jet launching from these
disks. Although we have not observed the launching of a jet, we did find
strong evidence for continued funnel clearing. We have identified the
characteristic funnel magnetization and clearing time scales. In
particular we found that the magnetization scale-height of the funnel and
the magnetic energy associated with it follow a power law, with the
dominant component being $\approx t^2$. Interestingly, by the end of our
simulation a region extending $140\, \rm km$ above the BH started to reach a strong
magnetization as indicated by the plasma beta parameter approaching
unity. Finally, we found that a continuous Poynting luminosity of
$10^{49}\, {\rm erg\,s^{-1}}$ is driven at late times.

Future work will be required to further investigate whether jet launching
is possible in these systems. Crucially, higher resolutions and longer
simulation times $>1\,\rm s$ will be needed. In that context it is of
particular relevance to point out that the parameter ranges found for
realistic accretion disks in terms of composition, disk mass, specific
entropy and magnetic-field topology can be used to initialize such
simulations, thereby removing the need to perform full numerical
relativity simulations to accurately study the long-term evolution of
this problem.

\section*{Acknowledgements}

ERM thanks Carolyn Raithel for helpful discussions. The authors thank
the anonymous referee for useful comments on disk self-regulation. ERM
gratefully acknowledges support from a joint fellowship at the Princeton
Center for Theoretical Science, the Princeton Gravity Initiative and the
Institute for Advanced Study. The simulations were performed on the
national supercomputer HPE Apollo Hawk at the High Performance Computing
Center Stuttgart (HLRS) under the grant numbers BBHDISKS and BNSMIC. The
authors gratefully acknowledge the Gauss Centre for Supercomputing
e.V. (www.gauss-centre.eu) for funding this project by providing
computing time on the GCS Supercomputer SuperMUC at Leibniz
Supercomputing Centre (www.lrz.de). LR gratefully acknowledges funding
from HGS-HIRe for FAIR; the LOEWE-Program in HIC for FAIR; ``PHAROS'',
COST Action CA16214.

\section*{Data availability}
Data is available upon reasonable request from the Corresponding Author.

\bibliographystyle{mnras}
\bibliography{aeireferences}

\appendix

\section{\texttt{BHB}$\Lambda\Phi$ models}
\label{sec:appendix_a}

\begin{figure*}
  \centering
  \includegraphics[width=\textwidth]{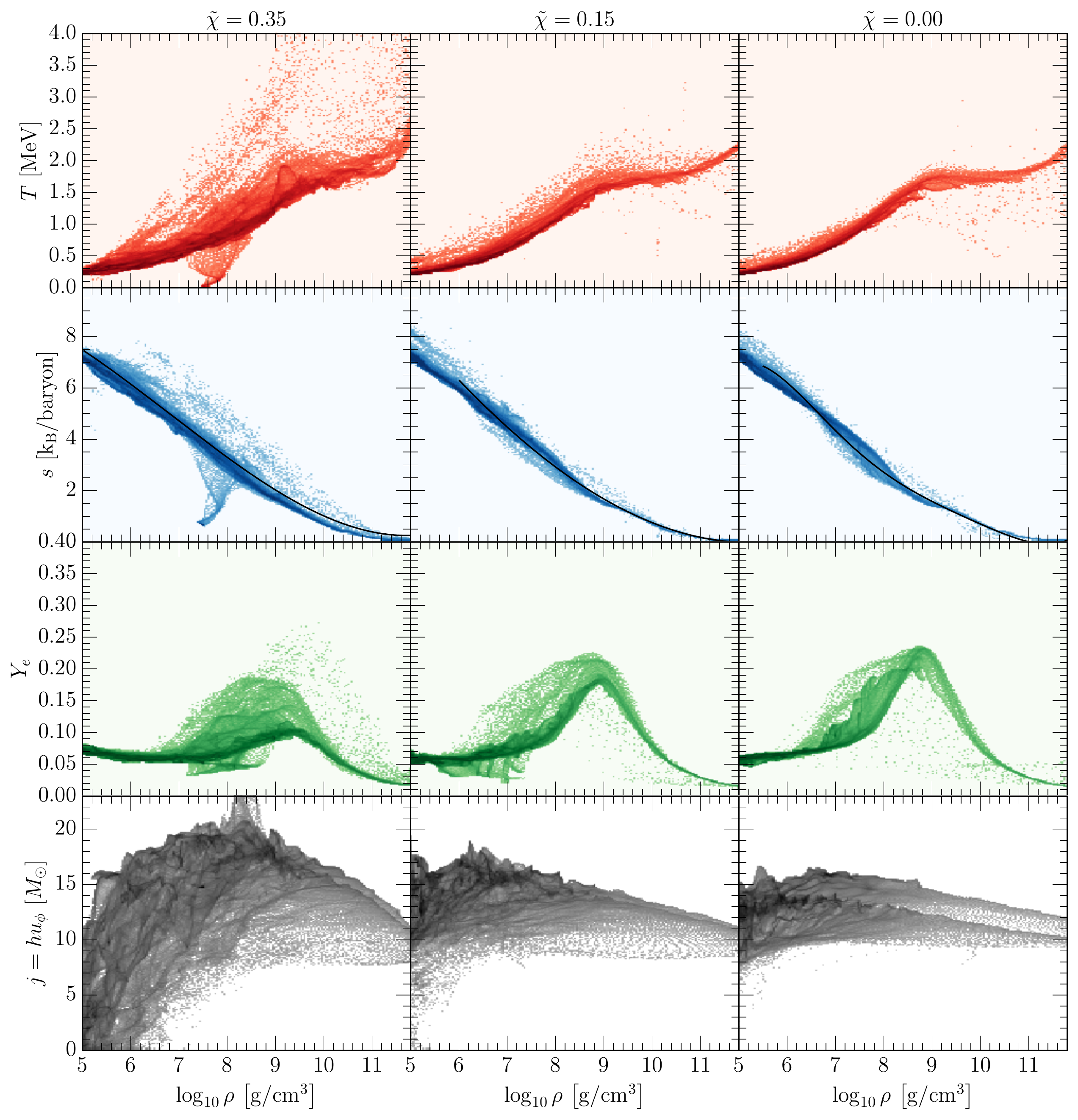}
  \caption{Same as Fig. \ref{fig:Dp_TNT} but for simulations using the
    \texttt{BHB}$\Lambda\Phi$.}
  \label{fig:Dp_BHB}
\end{figure*}

\begin{figure*}
  \centering
  \includegraphics[width=0.9\textwidth]{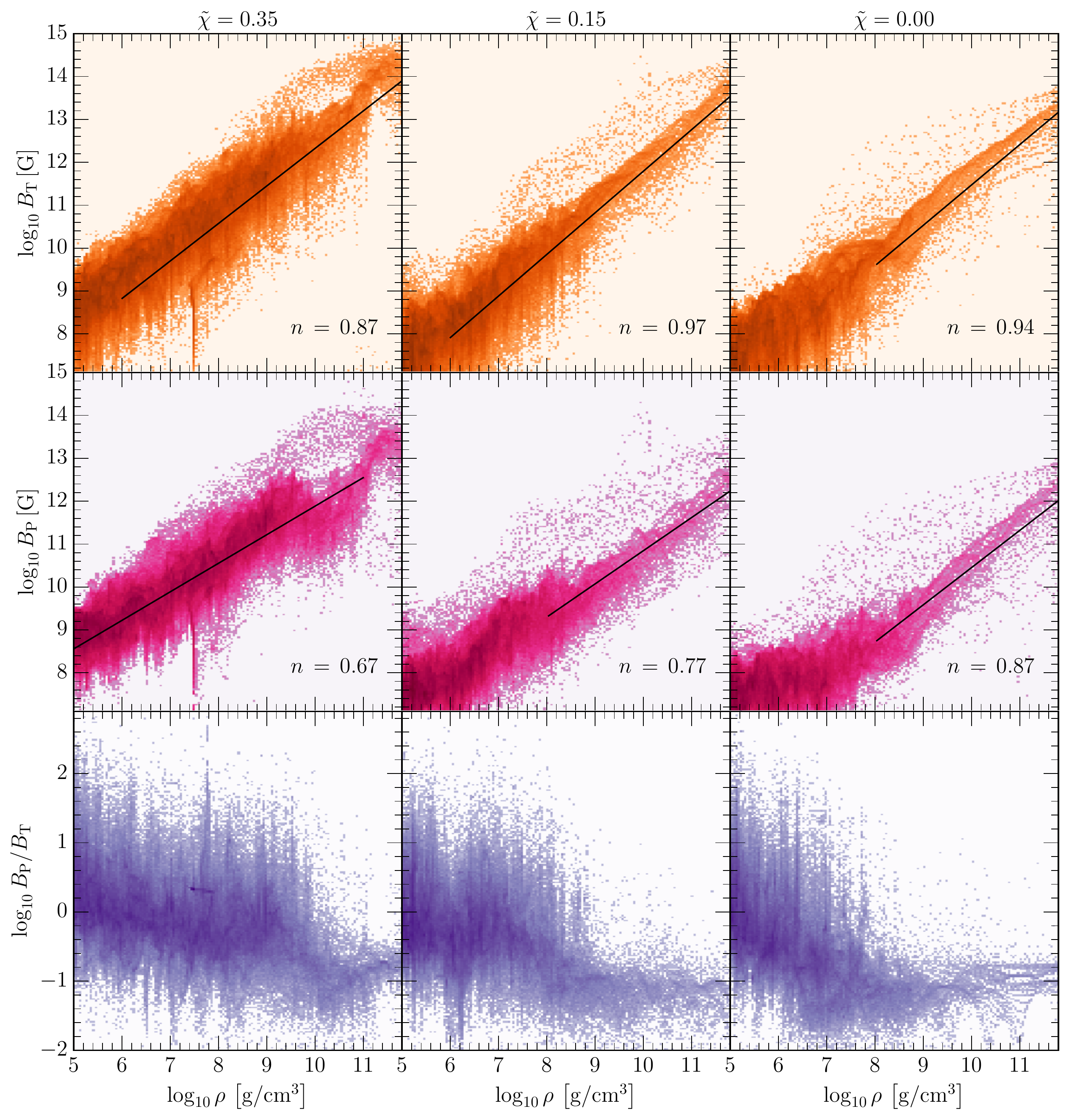}
  \caption{Same as Fig. \ref{fig:Bp_TNT} but using the
    \texttt{BHB}$\Lambda\Phi$ EOS.}
  \label{fig:Bp_BHB}
\end{figure*}

In this appendix, we show the disk properties for models using the
\texttt{BHB$\Lambda\Phi$} EOS. In particular, Fig. \ref{fig:Dp_BHB} shows
the hydrodynamical properties, whereas Fig. \ref{fig:Bp_BHB} shows the
topology of the magnetic field present in the disk. These figures should
be contrasted with the equivalent representations in
Figs. \ref{fig:Dp_TNT} and \ref{fig:Bp_TNT}, which refer instead to the
\texttt{TNTYST} EOS (see Secs. \ref{sec:hydro_prop} and \ref{sec:mag} for
a discussion of those results).

\end{document}